\documentclass[twocolumn]{aastex63}
\usepackage{xcolor}
\usepackage{gensymb}
\usepackage[nointegrals]{wasysym}
\usepackage{amsmath}

\usepackage{threeparttable}
\usepackage{bm}

\LTcapwidth=\textwidth

\shorttitle{}
\shortauthors{Kunimoto \& Matthews}

\begin{document}

\title{Searching the Entirety of \textit{Kepler} Data. II. Occurrence Rate Estimates for FGK Stars}

\correspondingauthor{Michelle Kunimoto}
\email{mkunimoto@phas.ubc.ca}

\author[0000-0001-9269-8060]{Michelle Kunimoto}
\affiliation{Department of Physics and Astronomy, University of British Columbia, 6224 Agricultural Road, Vancouver, BC V6T 1Z1, Canada}

\author[0000-0002-4461-080X]{Jaymie M. Matthews}
\affiliation{Department of Physics and Astronomy, University of British Columbia, 6224 Agricultural Road, Vancouver, BC V6T 1Z1, Canada}

\begin{abstract}
We present exoplanet occurrence rates estimated with approximate Bayesian computation for planets with radii between 0.5 and 16 $R_{\bigoplus}$ and orbital periods between 0.78 and 400 days, orbiting FGK dwarf stars. We base our results on an independent planet catalogue compiled from our search of all $\sim$200,000 stars observed over the \textit{Kepler} mission, with precise planetary radii supplemented by \textit{Gaia} DR2-incorporated stellar radii. We take into account detection and vetting efficiency, planet radius uncertainty, and reliability against transit-like noise signals in the data. By analyzing our FGK occurrence rates as well as those computed after separating F-, G-, and K-type stars, we explore dependencies on stellar effective temperature, planet radius, and orbital period. We reveal new characteristics of the photoevaporation-driven ``radius gap'' between $\sim$1.5 and 2 $R_{\bigoplus}$, indicating that the bimodal distribution previously revealed for $P < 100$ days exists only over a much narrower range of orbital periods, above which sub-Neptunes dominate and below which super-Earths dominate. Finally, we provide several estimates of the ``eta-Earth'' value --- the frequency of potentially habitable, rocky planets orbiting Sun-like stars. For planets with sizes $0.75 - 1.5$ $R_{\bigoplus}$ orbiting in a conservatively defined habitable zone ($0.99 - 1.70$ AU) around G-type stars, we place an upper limit (84.1th percentile) of $<0.18$ planets per star.
\end{abstract}

\section{Introduction}\label{sec:intro}

Determining the abundance of Earth-size planets in the habitable zones (HZs) of their stars, where liquid water could exist on a rocky planet's surface, is one of the major goals of exoplanetary science. Along with important implications for exoplanet habitability and prospects for extrasolar life, estimating this ``eta-Earth'' ($\eta_{\bigoplus}$) value informs the design of future missions focused on exoplanet detection and characterization. NASA's first exoplanet-finding mission, \textit{Kepler}, was specifically designed with this goal in mind \citep{bor11}. Aside from being the first (and so far only) mission capable of finding and characterizing Earth-sized planets in year-long orbits around Sun-like stars, \textit{Kepler} revolutionized our perspective on the diversity of planets in the Milky Way, having found more than half of all planets known today.\footnote{Based on exoplanet counts listed on the NASA Exoplanet Archive: https://exoplanetarchive.ipac.caltech.edu/\linebreak docs/counts\textunderscore detail.html}

However, calculating $\eta_{\bigoplus}$ is not straightforward. Finding Earth-size planets is challenging due to their small sizes and low transit signal-to-noise ratios (S/Ns), meaning planet detection pipelines have greater difficulty uncovering them than larger planets, and a higher risk of confusing them with transit-like noise in the data. Finding such planets in the HZs of Sun-like (G-dwarf) stars has added an difficulty due to their year-long orbits, necessitating a bare minimum of several years of observations to observe a few transits. Common definitions of the HZ also place potentially habitable planets hundreds of days outside even the $P \approx 500$ day sensitivity limit of \textit{Kepler}, necessitating the extrapolation of occurrence rates based on smaller orbital periods. There is also no standardized consensus yet on what defines the limits of the HZ, nor what range of planet sizes should be considered potentially habitable.

Furthermore, shortcomings affecting general planet occurrence rates include differences in accounting for imperfect detection efficiency (correcting for the ``search completeness'' of the planet sample), with previous works assuming an analytic function of the S/N \cite[e.g.][]{you11, how12}, empirically estimating detection efficiency by injecting synthetic planet transit signals into light curves and testing recovery \cite[e.g.][]{pet13, chr15}, or otherwise assuming that the catalogue is complete \cite[e.g.][]{cat11}. The efficiency of the vetting performed on detected signals may also be imperfect --- in the form of candidacy tests either incorrectly failing a planet as a false positive (FP), or passing a nonplanet signal as a planet --- though most previous studies have ignored this consideration from their estimates. In particular, incorporating an estimate of the ``reliability'' of a planet sample against transit-like noise was only performed for the first time in \citet{bry19}. Furthermore, nearly all previous studies have ignored uncertainty in planet radius, instead assuming that a detected planet's measured radius is exactly its true radius. This may have a significant effect on occurrence rates \citep{sha19}. Lastly, even when using the exact same dataset and characterization of completeness, different methods used to calculate occurrence rates can produce inconsistent results \cite[e.g. compare][]{pet13, for14}.

Reflecting these complications, $\eta_{\bigoplus}$ values in the literature span orders of magnitude. At one end, \citet{cat11} found an $\eta_{\bigoplus}$ between 0.01 and 0.03 planets per star; at the other, \citet{gar18} estimated an $\eta_{\bigoplus}$ greater than 1. Thus, new estimates are invaluable in bringing the exoplanet community toward consensus. This is the primary motivation behind our work.

Our approach to deriving exoplanet occurrence rates is largely inspired by the method first outlined in \citet{hsu18} and later expanded in \citet{hsu19}. \citet{hsu18} introduced using approximate Bayesian computation (ABC) as a tool to compute occurrence rates in a 2D grid of orbital period and planet radius. We will directly incorporate both search and vetting completeness using injection/recovery tests, as well as estimates of the reliability of our catalogue against transit-like noise. Furthermore, ABC is able to take into account uncertainty in planet radius, in contrast to the commonly-used grid-based inverse detection efficiency method (IDEM) which relies on knowing the planet properties exactly. Despite its popularity, IDEM has been shown to be less accurate than other methods and can produce artificially sharp features \citep{for14}, and may be especially biased toward lower rates near the detection limit \citep{hsu18}. IDEM is also unable to calculate occurrence rates over grid cells without planet detections, whereas ABC is able to place an upper limit. This is especially important for the $\eta_{\bigoplus}$ regime where planet detections are rare or even nonexistent depending on stellar sample cuts.

Our investigations will also allow us to comment on the greater period-radius space accessible by \textit{Kepler}. By splitting up our stellar sample over F-, G-, and K-type stars, we are able to investigate how planet occurrence rates change with stellar effective temperature; by calculating occurrence rates over a wide range of orbital periods ($P = 0.78 - 400$ days), we can comment on dependencies with period; and by calculating occurrence rates over a wide range of planet radii ($R_{p} = 0.5 - 16$ $R_{\bigoplus}$), we can comment on dependencies with radius. Similar population analyses have improved our understanding of planet formation and planet evolution \cite[e.g.][]{how12, fre13, ful17, pet18}. Finally, extrapolating our results to several different definitions of the $\eta_{\bigoplus}$ regime will provide estimates to consider alongside previous values in the literature.

\subsection{Paper Outline}

We describe our input stellar and planet catalogues in \S\ref{sec:inputs}. A full description of the process to create our planet catalogue is the content of \citet{kun20}, hereafter ``Paper I.'' In \S\ref{sec:comp}, we describe our determination of both search and vetting completeness using injection/recovery tests. In \S\ref{sec:method}, we give an overview of the ABC methodology, and we discuss our application of ABC to exoplanet occurrence rates in \S\ref{sec:application}. We make our code available for public use on Github\footnote{httops://github.com/mkunimoto/Exo-Occurrence} under the BSD 3-Clause License \citep{kun20code}.

Our overall results are presented in \S\ref{sec:results}, in which we discuss the dependence of exoplanet occurrence rates on stellar effective temperature (\S\ref{sec:temperature}), planet radius (\S\ref{sec:radius}), and orbital period (\S\ref{sec:period}). We also describe our incorporation of a simple catalogue reliability model to assess the impact of a nonzero FP rate and better constrain our estimates (\S\ref{sec:reliability}). In \S\ref{sec:etaearth}, we present both baseline and reliability-incorporated results over the potentially habitable, rocky exoplanet parameter space. Finally, in \S\ref{sec:limitations}, we review the limitations of our methodology and give a final recommended $\eta_{\bigoplus}$ estimate.

\section{Input Catalogues}\label{sec:inputs}
\subsection{Stellar Sample}
We started with the 197,096 \textit{Kepler} stars in the Q1-Q17 DR25 stellar catalogue \citep{mat17}, and calculated limb-darkening coefficients using $T_{\text{eff}}$, log$g$, and [Fe/H] from \citet{cla11}. With the arrival of stellar parallaxes in \textit{Gaia} Data Release 2 (DR2), \citet{ber18} produced improved radii for 177,911 targets, yielding an average radius precision of less than 10$\%$ for most \textit{Kepler} stars. Given that a fully updated set of stellar properties has not yet been released, we used these radii in tandem with the \citet{mat17} catalogue and only kept stars present in both catalogues.

We removed stars flagged in \citet{ber18} as likely binary stars (BIN flag = 1 or 3; 174,769 stars remained). We did not remove those flagged as binaries due to companions revealed with high-resolution imaging (BIN flag = 2), as these observations were only available for a subset of stars. Because the focus of our study is FGK dwarfs, we also removed stars with Evol flag $>$ 0 (116,637 stars remained), which indicate that they are unlikely to be on the main sequence.

To ensure each star's light curve had enough data to allow for the discovery of long-orbit planets, we required that the time length of the data ($T_\text{obs}$) is at least 2 yr, and the duty cycle ($f_{\text{duty}}$) was at least 0.6; in other words, at least 60$\%$ of the observations must be filled. After these cuts, 100,823 stars remained.

Lastly, we retained only FGK stars by using suggested $T_{\text{eff}}$ limits from \citet{pec13}. This left 40,010 F- ($6000 \leq T_{\text{eff}} < 7300 K$), 39,173 G- ($5300 \leq T_{\text{eff}} < 6000 K$), and 17,097 K-type ($3900 \leq T_{\text{eff}} < 5300 K$) stars, for a total of 96,280 stars in our sample.

Some stars in this sample may have been chosen as targets for reasons other than the \textit{Kepler} exoplanet search program, such as for asteroseismology. These stars would be expected to exhibit different noise and variability properties than typical main-sequence stars, which could introduce a systematic bias in our results relative to studies that focus on only exoplanet search targets. We checked the investigation ID of each star in our sample using the \textit{Kepler} Data Search \& Retrieval form on the Mikulski Archive for Space Telescopes (MAST)\footnote{https://archive.stsci.edu/kepler/data\textunderscore search/search.php}, and found that 290 did not have an ``EX*'' ID. In other words, 99.7\% of our 96,280 FGK stars were selected for the exoplanet search program, and we do not expect a significant bias to be present.

\subsection{Planet Sample}

Our full search and vetting pipeline is described in Paper I. In short, we obtained Q1-Q17 DR25 long-cadence PDC light curves from the MAST. We detrended each light curve using the \texttt{detrend5} routine from the Kepler Transit Model Codebase \citep{row16}, where each observation was corrected by fitting a cubic polynomial to a segment (typically two days wide) centred on the time of measurement. We then 5$\sigma$-clipped the data, removing outliers only in the positive flux direction so as to leave deep transits untouched and removed data near data gaps. Then, we used a \citet{kov02} box least-squares (BLS) algorithm to search for potential transits. After identifying an event in the light curve, we calculated its S/N by dividing the mean transit depth by the standard error of the mean, giving

\begin{equation}\label{eqn:S/N0}
    \text{S/N} = \frac{\sqrt{N}}{\sigma} T_{\text{dep}}
\end{equation}

\noindent where $T_{\text{dep}}$ is the mean transit depth, $\sigma$ is the standard deviation of the observations, and $N$ is the number of in-transit data points. This definition is comparable to the ``effective'' S/N described in \citet{kov02}. Following \citet{row14}, we estimated $\sigma$ using the standard deviation of all out-of-transit observations --- defined as data outside of two transit durations of the centre of the detected signal --- and used the median absolute deviation (MAD) with $\sigma = 1.48$MAD \citep{hoa83} to be more robust to outliers.

To define transit candidates (TCs), we followed the suggestion of \citet{kov02} that the threshold for a significant detection with the BLS algorithm is S/N = 6. We also required at least three transits and the passing of an initial vetting stage to reject false alarms caused by instrumental and astrophysical systematics. 

Each TC was passed through a vetting pipeline, involving both machine and manual triage. Automated candidacy tests were used to flag both noise false alarms and astrophysical FPs, while visual inspection was used as a ``reality check'' to confirm each surviving TC as a planet candidate (PC).

Around the 96,280 FGK stars considered, we identified 2623 PCs that matched with already known planet candidates as listed on the NASA Exoplanet Archive\footnote{https://exoplanetarchive.ipac.caltech.edu/, accessed 2019 May 9}, defined as \textit{Kepler} Objects of Interest (KOIs) with either a CONFIRMED or CANDIDATE disposition. Additionally, we introduce eight previously unknown candidates from our Paper I search, for a total of 2631 planets in our full catalogue. By comparison, \textit{Kepler}'s Q1-Q17 DR25 pipeline identified 2829 planet candidate KOIs corresponding to this stellar sample.

\subsubsection{Confirmed and Candidate KOIs Missed}

We had a 98.9$\%$ recovery rate for all confirmed FGK KOIs, finding and passing 1655 of 1673. Nine of the KOIs (KOI-172.02, 701.04, 1236.03, 2038.03, 2365.02, 4034.01, 4384.01, 5706.01, and 7016.01) were either very close to passing the vetting pipeline, or failed only one of our tests, while four (KOI-245.03, 490.02, 1274.01, 3234.01) were detected but failed to meet the requirements to become a TC. KOI-490.02 and KOI-1274.01 were strong signals, but had less than the required three transits. The only planet completely missed was KOI-245.04, though we note that despite its Confirmed Exoplanet Archive Disposition, it is also flagged as a Not Transit-Like FP.

Another four of the failed confirmed KOIs (KOI-142.01, 377.01, 377.02, and 884.02) displayed significant transit timing variations (TTVs). Because our vetting pipeline did not correct for TTVs, it is unsurprising that these failed despite their high S/N. Given the unique nature of these planets and considering that neither our search nor vetting completeness models take into account TTVs, we decided to include these in our catalogue.

We summarize all confirmed planets not included in our catalogue in Table \ref{tbl:missed}. We note that four were also missed by the the Q1-Q17 DR25 pipeline, and six that were detected may not necessarily be considered ``high-quality'' candidates \cite[e.g. requiring Disposition Score $> 0.9$; ][]{mul18}.

We had a much lower recovery rate of candidate KOIs, finding 961 of 1487 (64.6$\%$). A lower rate is to be expected considering that confirmed planets typically have higher S/N and transit shapes more clearly consistent with a planetary origin. Furthermore, 299 (around $60\%$) of the candidates missed or failed by our pipeline were not detected by the DR25 pipeline.

Our goal was to produce an independent pipeline that could both search for planets and be used for completeness modeling conducive to occurrence rate statistics. Thus, with the exception of the confirmed KOIs failed due to exhibiting TTVs, we do not include any of the the KOIs missed or failed in our determination of occurrence rates.

\begin{table*}[ht!]
\centering
\caption{Confirmed planet KOIs corresponding to the FGK stars in our sample missed or failed by our pipeline. Table entries are taken from the NASA Exoplanet Archive.}\label{tbl:missed}
\begin{tabular}{c|c|c|c|c|c}
\hline\hline
    KOI & $P$ (days) & $R_{p}$ ($R_{\bigoplus}$) & S/N & Disposition Score & TCE Delivery \\
\hline
    172.02 & 242.5 & 1.73 & 23.20 & 0.6930 & Q1-Q17 DR25 \\
    245.03 & 13.4 & 0.27 & 7.40 & - & - \\
    245.04 & 51.2 & - & - & - & - \\
    490.02 & 1071.2 & 9.27 & 544.20 & 0.0000 & Q1-Q17 DR25 \\
    701.04 & 267.3 & 1.43 & 19.30 & 0.0000 & Q1-Q17 DR25 \\
    1236.03 & 54.4 & 3.20 & 44.90 & - & Q1-Q17 DR24 \\
    1274.01 & 705.0 & 4.53 & 96.10 & - & - \\
    2038.03 & 17.9 & 1.39 & 11.40 & 0.8890 & Q1-Q17 DR25 \\
    3234.01 & 2.4 & 0.85 & 13.40 & 0.9930 & Q1-Q17 DR25 \\
    4034.01 & 7.0 & 6.14 & 18.60 & 0.1000 & Q1-Q17 DR25 \\
    4384.01 & 122.4 & 2.15 & 12.20 & 0.9970 & Q1-Q17 DR25 \\
    5706.01 & 425.5 & 3.20 & 19.60 & 0.9040 & Q1-Q17 DR25 \\
    7016.01 & 384.8 & 1.09 & 12.30 & 0.7710 & Q1-Q17 DR25 \\ 
\end{tabular}
\end{table*}

\subsubsection{New PCs}

We added eight new candidates to our FGK planet catalogue, listed in Table \ref{tbl:newcands}. As discussed in Section 6 of Paper I, these candidates passed our full vetting pipeline, and underwent additional analysis including astrophysical FP probability (FPP) calculation. We used \texttt{vespa}, a Python package that assesses the likelihood that a transit signal is caused by a planet compared to astrophysical scenarios such as grazing or background eclipsing binaries \citep{mor12,mor18}. \texttt{vespa} has already been used to validate over a thousand KOIs \citep{mor16} using a threshold of FPP $< 0.01$, which six of our eight candidates met. 

\begin{table}[ht!]
\centering
\caption{New planet candidates added to our FGK planet catalogue from Paper I. Candidates are listed according to their \textit{Kepler} Input Catalogue (KIC) ID.}\label{tbl:newcands}
\begin{tabular}{c|c|c|c}
\hline\hline
    KIC & KOI & $P$ (days) & $R_{p}$ ($R_{\bigoplus}$)\\
    \hline
    2696784 b & - & 82.3 & 1.50 \\
    2861140 b & - & 36.9 & 2.28 \\
    6126245 b & - & 3.5 & 0.68 \\
    6782399 b & - & 34.2 & 1.65 \\
    7747788 b &-  & 133.1 & 1.67 \\
    11350118 c & 4509.02 & 2.7 & 0.66 \\
    11805835 b &-  & 23.5 & 0.94 \\
    12023559 b &-  & 84.6 & 1.86 \\
\end{tabular}
\end{table}

\subsubsection{Planet Properties}

As part of the vetting pipeline, we found a least-squares best fit of each planet transit with a \citet{man02} quadratic limb-darkening transit model assuming circular orbits. The model is parameterized by orbital period ($P$), transit epoch ($T_{0}$), ratio of the planet and star radii ($R_{p}/R_{s}$), distance between planet and star at midtransit in units of stellar radius ($a/R_{s}$), impact parameter ($b$), and zero-point flux ($z$). Following the vetting pipeline, we refit each transit using \texttt{emcee}, a Python implementation of an affine invariant Markov Chain Monte Carlo (MCMC) ensemble sampler \citet{for13}. We set $P$ and $T_{0}$ fixed to their least-squares values to aid in convergence and initialized 50 walkers in a tight Gaussian ball centred on the rest of the least-squares best-fit parameters. We ran the sampler for up to 100,000 steps per walker and checked the autocorrelation time every 100 steps. We considered the algorithm converged if the chain was longer than 100 times the estimated autocorrelation time and if the estimate changed by less than 1\% from the previous estimate. Ninety-nine percent of the chains converged under this criteria. For burn-in, we removed the first number of steps from each chain equal to twice the autocorrelation time.

For the 1\% of planets that did not converge, we reverted to their least-squares best-fit values. Given that this study is primarily a population analysis, and only the orbital period and a ratio of the planet to star radius are needed for each model fit, detailed analysis of each system is beyond the scope of this study. We repeated our analysis excluding the planets that did not converge and found that median occurrence rates did not change by more than 6\%, and all variations were well within 1$\sigma$ uncertainty.

For the planets with TTVs, we used the fit results listed on the NASA Exoplanet Archive.

\subsubsection{Dilution}

The planet radius $R_{p}$ can be determined from the fitted parameter $R_{p}/R_{s}$ by multiplying by the known stellar radius. However, there may be one or more nearby stars that contribute light to the \textit{Kepler} aperture, causing the measured transit depth to be diluted. In these cases, the planet radius can be underestimated. We make the assumption that the planet orbits the brighter primary star, in which case we apply a correction factor to the planet radius in the form of

\begin{equation}\label{corr1}
    R_{p,\text{corr}} = R_{p}\sqrt{1 + 10^{-0.4\Delta m}},
\end{equation}

\noindent where $\Delta m = m_{\text{sec}} - m_{\text{pri}}$ is the \textit{Kepler} magnitude difference between the primary and secondary star. For more than one companion per star, the previous equation becomes

\begin{equation}\label{corr2}
    R_{p,\text{corr}} = R_{p}\sqrt{1 + \sum_{i=1}^{N} 10^{-0.4\Delta m_{i}}},
\end{equation}

\noindent where the sum is for $N$ companion stars with magnitude differences $\Delta m_{i}$.

We used the high-resolution imaging results from the \textit{Kepler} Follow-Up Observation Program \citep{fur17} to correct the radii of planets around stars with a potential companion within 4$^{\prime\prime}$, the size of a \textit{Kepler} pixel. \citet{fur17} compiled observations for a total of 3557 KOIs, including those observed in the first three Robo-AO surveys \citep{law14, bar16, zie17}, and provided a weighted average of correction factors across a variety of bands for 1891 KOIs with companions. 

\citet{zie18} presented a fourth Robo-AO survey for 532 KOIs published after \citet{fur17}. Their results were provided as $\Delta m$ in the LP600 band, which we approximate to be equal to the \textit{Kepler} band for use in Eqns. \ref{corr1} and \ref{corr2}. We also used our own adaptive optics imaging follow-up described in Paper I for three of our new FGK PCs (KIC-6126245 b, 6782399 b, and 7747788 b), none of which had a nearby stellar companion.

In total, 2578 of the 2631 PCs (98.0$\%$) in our FGK sample had high-resolution imaging observations, and we applied correction factors to 679.


\subsubsection{Final Planet Catalogue}

\begin{figure}[ht!]
\centering
\includegraphics[width=0.5\textwidth]{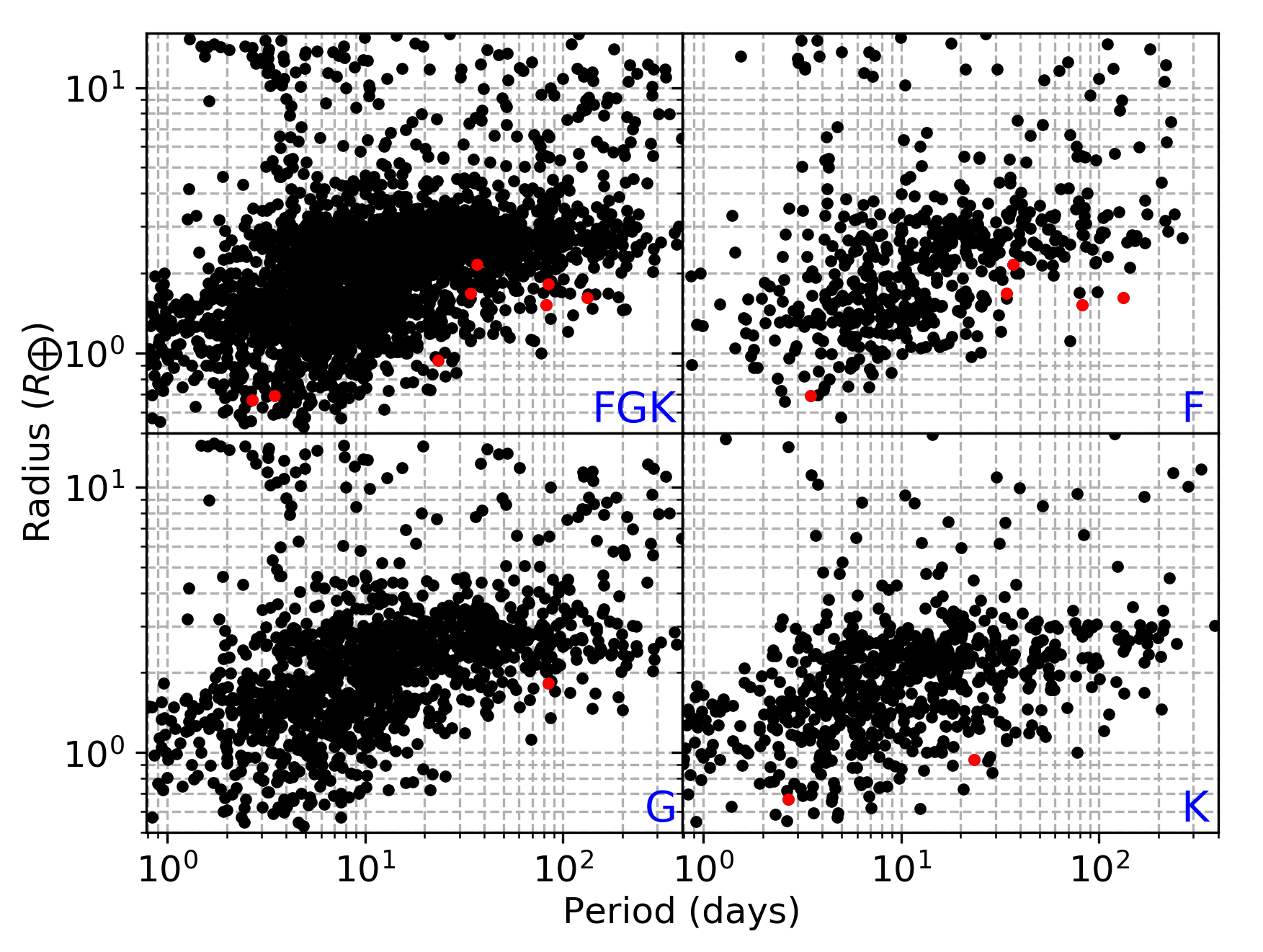}
\caption{Planets in our final catalogue, plotted according to orbital period and radius. Plots are organized by host star stellar type, including F- ($6000 \leq T_{\text{eff}} < 7300 K$), G- ($5300 \leq T_{\text{eff}} < 6000 K$), and K-type ($3900 \leq T_{\text{eff}} < 5300 K$) stars. Planets new to our work (from Paper I) are plotted in red.}\label{fig:FGKplanets}
\end{figure}

Our focus for this paper is on the occurrence rates of planets in a period-radius grid spanning orbital periods $0.78125 < P < 400$ days and radii $0.5 < R_{p} < 16.0$ $R_{\bigoplus}$. Lower and upper limits on these properties were chosen so as to split the grid into logarithmically spaced bins comparable to bins used in previous grid-based works \cite[e.g.][]{how12, pet13, mul15a}. After applying the radius correction factors and including only candidates that fit these criteria, our final planet catalogue involved 557 candidates around F-type stars, 1276 around G-type stars, and 700 around K-type stars, for a total of 2533 planet candidates. Table \ref{tbl:specnum} summarizes the sizes of each star and planet sample, while Fig. \ref{fig:FGKplanets} shows the distribution of planets based on orbital period and radius.


\begin{table}[ht!]
\centering
\caption{Number of stars and planets by stellar type. $N_{\text{DR25}}$ gives the number of planets found in DR25 around the same sample of stars for comparison.}\label{tbl:specnum}
\begin{tabular}{c|c|c|c|c|c}
\hline\hline
    Type & $T_{\text{min}}$ ($K$) & $T_{\text{max}}$ ($K$) & $N_{\text{stars}}$ & $N_{\text{planets}}$ & $N_{\text{DR25}}$\\
\hline
    FGK & 3900 & 7300 & 96,280 & 2,533 & 2,700\\
    F & 6000 & 7300 & 40,010 & 557 & 639\\
    G & 5300 & 6000 & 39,173 & 1,276 & 1,338\\
    K & 3900 & 5300 & 17,097 & 700 & 723\\
    \end{tabular}
\end{table}

\section{Completeness Model}\label{sec:comp}

Our planet sample is not expected to be ``complete/'' Particularly near the detection limit, transiting planets are often missed or even mislabeled as FPs. Thus, it is important to quantify the completeness corrections for both our transit detection pipeline and vetting pipeline to derive accurate occurrence rates. Here, search completeness refers to the fraction of transiting planets that are detected, while vetting completeness refers to the fraction of detected planets that are correctly classified as planet candidates. 

Search completeness is a common feature of occurrence rate studies, and is typically estimated by assuming an analytic function of the S/N \cite[e.g.][]{you11, how12} or by injecting synthetic planet transit signals into light curves and testing recovery \cite[e.g.][]{pet13, chr15}. For instance, using injection/recovery tests, \citet{chr15} showed that the \textit{Kepler} detection efficiency is well modeled by a gamma cumulative distribution function, of the form 

\begin{equation}\label{eqn:Pdet}
    P_{\text{det}}(\text{S/N}) = \frac{c}{b^{a}(a -1)!}\int_{0}^{\text{S/N}}x^{a-1} e^{-x/b} dx,
\end{equation}

\noindent giving the probability of detecting a transit with a given signal-to-noise ratio S/N.

However, vetting completeness has often been ignored, with most previous studies assuming perfect efficiency at classifying planet transit signals as planets. In a comparison between \textit{Kepler} DR25 occurrence rates derived under this assumption and various vetting models, \citet{hsu19} found that taking into account imperfect vetting was important for small planets ($R_{p} < 2 R_{\bigoplus}$) and planets with orbital periods longer than a month ($P > 32$ days). They also found that their occurrence rates were robust to the choice of vetting model, as differences between the two models tested were still significantly smaller than the uncertainty due to the \textit{Kepler} sample size. 

With these considerations, we adopted the \citet{hsu19} combined detection and vetting efficiency model described in Section 2.2.2 of their paper, using injection/recovery tests to determine the fraction of planets both successfully detected and vetted by the automated pipeline. These results were fit to the \citet{chr15} gamma cumulative distribution function, and a direct dependence on the number of transits $N_{\text{tr}}$ is introduced by fitting separate functions for injections with 3, 4, 5, 6, 7-9, 10-18, 19-36, and $\geq$ 37 transits.

Similar to \citet{pet13}, we injected 96,280 planet transits, one for each FGK star in our sample, into Q1-Q17 light curves downloaded from the MAST. Half the signals were log-uniformly distributed over $0.78 < P < 100$ days and $0.5 < R_{p} < 16.0$ $R_{\bigoplus}$, with the other half log-uniformly distributed over $100 < P < 500$ days so as to improve the determination of completeness for planets with low numbers of transits. Each transit was created using a quadratic limb-darkening \citet{man02} model, with impact parameters ($b$) uniformly distributed between 0 and 1 and circular orbits assumed.

We prepared, searched, and vetted the simulated data with the same process as for the actual observed data, using the federation process described in \citet{mul15} to match detections with the injected planets. The only exception was that we did not perform the manual vetting stage given that it would be infeasible to review the tens of thousands of simulated PCs that were passed by the automated stage. Thus, we assumed that the manual component is completely accurate at classifying planets. Using similar injection/recovery tests in Paper I, we estimated that the manual inspection would lower our overall vetting completeness by $\sim$1-2$\%$, which would indicate that this assumption should not significantly impact our occurrence rates.

Eqn. \ref{eqn:Pdet} requires an estimate of each injected transit's S/N as defined in Eqn. \ref{eqn:S/N0}, which we can find with the planet's known radius, period, and impact parameter, and basic properties known about the star and corresponding light curve.

First, we estimate the number of transits from the length of observations in the light curve and the planet's orbital period, taking into account loss of data with $f_{\text{duty}}$,

\begin{equation}
    N_{\text{tr}} = \frac{T_{\text{obs}}f_{\text{duty}}}{P}.
\end{equation}

\noindent The duration of the transit $T_{\text{dur}}$ can be estimated as

\begin{equation}\label{eqn:tdur}
    T_{\text{dur}} = \frac{R_{s}}{a}\frac{P}{\pi}\sqrt{1 - b^{2}},
\end{equation}

\noindent where $a$ is the semi-major axis of the orbit, from

\begin{equation}
    a^{3} = \frac{GM_{s}P^{2}}{4\pi^{2}}
\end{equation}

\noindent with stellar mass $M_{s}$. Combined with $N_{\text{tr}}$ and a rate of one observation every 29.42 minutes (one \textit{Kepler} long cadence), we estimate the total number of data points during transit as

\begin{equation}
    N = \frac{N_{\text{tr}}T_{\text{dur}}}{\text{29.42 min}}.
\end{equation}

\noindent Lastly, we calculate the expected depth of the transit $\delta$ from the ratio of planet to star radii, $k = R_{p}/R_{s}$, taking into account quadratic limb-darkening coefficients $u_{1}$ and $u_{2}$. \citet{zin19a} estimated this as

\begin{equation}
\begin{split}
    A & = 1 - (u_{1} + u_{2}) \\
    B & = \frac{A}{4} + \frac{u_{1} + 2u_{2}}{6} - \frac{u_{2}}{8} \\
    \delta & = 1 -\frac{1}{B}\bigg(\frac{A}{4} + \frac{(u_{1} + 2u_{2})(1 - k^{2})^{3/2}}{6} \\
    & - \frac{u_{2}(1 - k^{2})}{8}\bigg).
\end{split}
\end{equation}

\noindent Putting everything together, the expected S/N is

\begin{equation}\label{eqn:S/N}
\text{S/N} = \sqrt{N}\frac{\delta}{\sigma}
\end{equation}

\noindent where $\sigma$ is estimated using the MAD of the light curve with $\sigma = 1.48\text{MAD}$ \citep{hoa83}.

\begin{figure}[h!]
\centering
\includegraphics[width=\linewidth]{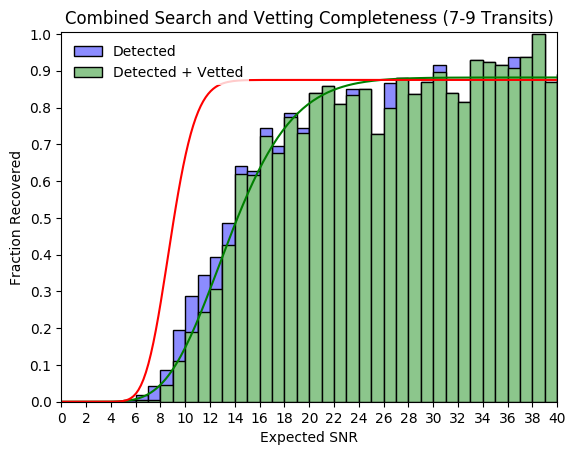}
\includegraphics[width=\linewidth]{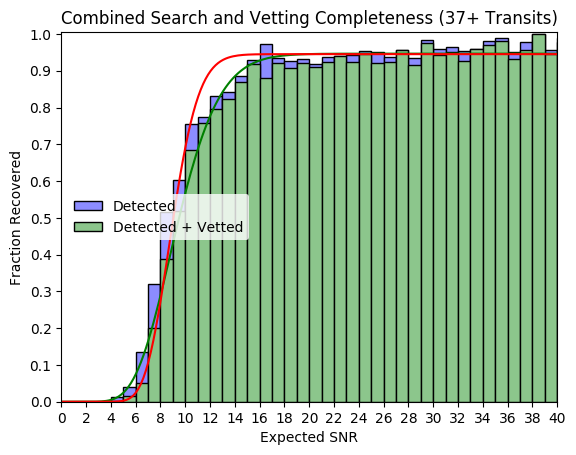}
\caption{Combined search and vetting completeness of our pipeline, showing the fraction of injected transits recovered based only on the search (blue) and both search and vetting (green). A gamma cumulative distribution function (Eqn. \ref{eqn:Pdet} is fit to the combined recovery fraction (green line). These examples corresponds to $N_{\text{tr}} = 7 - 9$ (top) and $N_{\text{tr}} \geq 37$ (bottom). For comparison, the \citet{hsu19} best-fit gamma CDF fits as a function of expected MES statistic are also shown (red lines).}\label{fig:comp}
\end{figure}

Fig. \ref{fig:comp} shows the fraction of successful detections as a function of expected S/N for $N_{\text{tr}} = 7 - 9$ and $N_{\text{tr}} \geq 37$ as examples. The recovery fractions based on the search pipeline alone and the combined search and vetting pipeline are shown for comparison. As expected, the vetting process affects recovery at lower S/N ($\leq 15$) significantly more than at higher S/N, and overall recovery is improved for planets with more transits. Our full fit results are shown in Table \ref{tbl:comp}.

Fig. \ref{fig:comp} and Table \ref{tbl:comp} also give the corresponding combined search and vetting completeness models from \citet{hsu19}. As a reminder, these were based on the \textit{Kepler} DR25 pipeline's injection/recovery tests \citep{chr17}. While our pipelines differ in how we define S/N (with our pipeline using the BLS S/N \citep{kov02}, and the \textit{Kepler} team using the so-called Multiple Event Statistic \cite[MES,][]{jen02}), they may be considered comparable, and we can comment on key differences in pipeline performance. In particular, the DR25 pipeline is significantly better at recovering low-S/N events and those with few transits. This is expected given our more simplistic pre-search data reduction. At higher S/Ns, especially for events with more transits, pipeline performance is more similar.

\begin{table}[h!]
    \centering
    \caption{Best-fit parameters for $P_{\text{det}}$, the combined search and vetting model, with comparisons to the DR25 model results of \citet{hsu19}.}
    \begin{tabular}{c|c|c|c|c|c|c}
    \hline
    \hline
    & \multicolumn{3}{c|}{This work} & \multicolumn{3}{c}{\citet{hsu19}}\\
    \hline
    $N_{\text{tr}}$ & $a$ & $b$ & $c$ & $a$ & $b$ & $c$\\
    \hline
    3 & 12.0239 & 1.3892 & 0.4653 & 33.3884 & 0.2645 & 0.6991\\
    4 & 17.4744 & 0.9059 & 0.6651 & 32.8860 & 0.2696 & 0.7684\\
    5 & 13.5488 & 1.0900 & 0.7704 & 31.5196 & 0.2827 & 0.8337\\
    6 & 11.4812 & 1.2763 & 0.8369 & 30.9919 & 0.2870 & 0.8599\\
    7-9 & 11.5413 & 1.2063 & 0.8817 & 30.1906 & 0.2947 & 0.8750\\
    10-18 & 11.4538 & 1.0725 & 0.9118 & 31.6432 & 0.2794 & 0.8861\\
    19-36 & 14.8651 & 0.7292 & 0.9164 & 32.6448 & 0.2689 & 0.8897\\
    $\geq$ 37 & 12.2332 & 0.7820 & 0.9465 & 27.8185 & 0.3243 & 0.9451\\
\end{tabular}
    \label{tbl:comp}
\end{table}

\section{Occurrence Rate Methodology}\label{sec:method}

\subsection{Approximate Bayesian Computation}\label{sec:abc}

Bayesian inference is an increasingly popular approach of statistical inference on unknown parameters. In this framework, Bayes' theorem is used to estimate the posterior probability distribution $P(\boldsymbol{\theta}|D)$ of a model with parameters $\boldsymbol{\theta}$ given the data $D$,

\begin{equation}
    P(\theta | D) = \frac{P(D|\theta)P(\theta)}{P(D)},
\end{equation}

\noindent where $P(D|\boldsymbol{\theta})$ is the likelihood function, indicating the compatibility of the data given the model; 
$P(\boldsymbol{\theta})$ is the prior probability, representing initial beliefs toward the model; and $P(D)$ is a normalization constant. The best-fit model parameters can be estimated from $P(\boldsymbol{\theta}|D)$ such as by finding the posterior mode (most probable values of $\boldsymbol{\theta}$) or posterior median (50th percentile), with credible intervals representing our uncertainty about the model parameters.

For simple models, the likelihood function can typically be derived analytically. However, for more complex models, the likelihood may be unknown or too computationally expensive to evaluate. It is in these cases that the ``likelihood-free'' method of ABC steps in as an effective and rigorous way of performing an approximate Bayesian analysis.

ABC circumvents the need for a likelihood function by using our prior information along with an ability to simulate, or ``forward model,'' the observed data under investigation. By simulating a large number of datasets and quantifying the ``distance'' between each dataset and the observed dataset, the distribution of model parameters that provides the best matches can be determined. This distribution serves as an approximation to the posterior probability distribution. 

\subsection{Population Monte Carlo ABC}

The specific form of ABC used here is the population MC (ABC-PMC) algorithm proposed by \citet{bea09}, wherein multiple generations of simulated data are created and an adaptive importance sampling scheme is used to evolve the ABC posterior. We use the ABC-PMC algorithm implemented in \texttt{cosmoabc}, a Python ABC Sampler \citep{ish15}, which is summarized here.

To initialize the ABC-PMC algorithm, we draw a set of $M$ values from the prior distribution, called ``particles,'' $\{\boldsymbol{\theta}^{i}\}$ with $i \in [1,M]$. $M$ is chosen to be much larger than $N$, the number of samples needed to characterize the prior. For each particle, we generate a simulated dataset $D_{S}^{i}$ and use a distance function $\rho$ to calculate the distance between the simulated and real dataset, $\rho^{i} = \rho(D, D_{S}^{i})$. From the whole set of $M$ particles, we keep only the $N$ particles with the smallest $\rho^{i}$. These constitute the zeroth ``generation'' ($S_{t=0}$), and the 75$\%$ quantile of all $\rho \in S_{t=0}$ gives the distance threshold for the next iteration ($\epsilon_{t=1}$). Each particle is assigned an equal weight, $W_{t=0}^{j} = 1/N$, for $j \in [1,N]$.

An importance sampling technique is used to produce subsequent generations ($t > 0$). We  draw a trial particle $(\boldsymbol{\theta}_{\text{try}})$ from the previous generation $S_{t-1}$ with weights $W_{t-1}$, and use it to simulate a catalogue and find its associated distance, $\rho_{\text{try}}$. We store $\boldsymbol{\theta}_{\text{try}}$ to the current generation $S_{t}$ if $\rho_{\text{try}} \leq \epsilon_{t}$. This process is repeated until $S_{t}$ is filled with $N$ accepted particles. We then calculate the weights of each particle as

\begin{equation}
    W_{t}^{j} = \frac{P(\boldsymbol{\theta}_{t}^{j})}{\sum_{i=1}^{N} W_{t-1}^{i} N(\boldsymbol{\theta}_{t}^{j}; \boldsymbol{\theta}_{t-1}^{i}, C_{t-1})}
\end{equation}

\noindent where $P(\boldsymbol{\theta}_{t}^{j})$ is the prior probability distribution calculated at $\boldsymbol{\theta}_{t}^{j}$, and $N(\boldsymbol{\theta}_{t}^{j}; \boldsymbol{\theta}_{t-1}^{i}, C_{t-1})$ represents a Gaussian probability density function (PDF) centred at $\boldsymbol{\theta}_{t-1}^{i}$ with covariance matrix built from $S_{t-1}$ and calculated at $\boldsymbol{\theta}^j$.

Following the determination of the new weights, the algorithm repeatedly produces new generations until subsequent iterations no longer significantly change the ABC posterior. In \texttt{cosmoabc}, this convergence occurs when the number of draws necessary to construct a generation is much larger than $N$.

\section{ABC Applied to Exoplanet Occurrence Rates}\label{sec:application}

Planet surveys have a variety of complexities that make the determination of the correct likelihood impractical, such as the existence of selection effects that are pipeline dependent, the choice of targets, and the measurement uncertainties in the planet properties. Thus, ABC is well suited to the inference of occurrence rates based on \textit{Kepler} planet catalogues and our independent catalogue outlined in Paper I.

As discussed in \S\ref{sec:method}, ABC depends on the following elements:

\begin{itemize}
    \item A prior probability distribution over the model parameters,
    \item A forward model, to simulate the data given the model parameters, and
    \item A distance function, to assess the agreement between the simulated data and the observed data.
\end{itemize}

Because we calculate occurrence rates over a 2D grid of orbital period and planet size in this work, the model parameters of interest are $f_{p,r}$, the average number of planets per star in period bin $p$ and radius bin $r$. We assume that each $f_{p,r}$ is constant over the relevant range of periods and radii. Meanwhile, the forward model must simulate the planet population around the considered stellar sample using each bin's guess occurrence rate and take into account selection effects and biases such as catalogue completeness and planet radius uncertainty to produce a simulated catalogue. The distance function must then compare the simulated catalogue to the actual observed catalogue to indicate which occurrence rates most closely describe the distribution.

\subsection{Prior Probability}

We assign independent uniform priors for each occurrence rate over $[0,f_{\text{max},p,r})$. The upper limit for each bin is

\begin{equation}
    f_{\text{max},p,r} = C \times \log_{2}\bigg(\frac{P_{\text{max},p}}{P_{\text{min},p}}\bigg)\times\log_{2}\bigg(\frac{R_{p,\text{max},r}}{R_{p,\text{min},r}}\bigg)
\end{equation}

\noindent with $C = 2$, small enough that proposals with more than three planets per factor of 2 in period are rare \citep{hsu19}. This is consistent with expectations based on long-term orbital stability.

\subsection{Forward Model}\label{sec:forwardmodel}

It is within our exoplanet population simulator that many of the complexities that make a likelihood function infeasible to compute are able to be incorporated into the determination of occurrence rates.

One such complexity is the existence of selection effects. \citet{you11} outlined three main selection effects to be accounted for as part of robust exoplanet population analysis. These are quantified as detection efficiencies, $\eta$, which give the ratio of detections to actual planets: (i) $\eta_{\text{tr}}$, the transit probability that the planet crosses our line of sight to the star; (ii) $\eta_{\text{rec}}$, the efficiency at which the detection pipeline recovers the planet; and (iii) $\eta_{\text{fp}} = 1/(1 - r_{\text{fp}})$, where $r_{\text{fp}}$ is the rate of FP events that are detected as planets. The net detection efficiency of a given planet is found by multiplying all of the above efficiencies together. For our baseline results, we assumed the FP rate is low enough that it can be ignored for simplicity ($\eta_{\text{fp}} = 1$). However, we discuss potential implications of this assumption in \S\ref{sec:reliability}.

Importantly, these selection effects change on a per-star basis. For instance, our completeness model depends on both the physical properties of a star and the characteristics of its associated \textit{Kepler} light curve. Our forward model allows us to take these into account and find a specific completeness for a planet around a specific star, with little sacrifice of computational efficiency. By comparison, studies that have used likelihood functions in occurrence rate statistics such as \citet{bur15} and \citet{zin19a} have had to utilize star-averaged detection efficiencies that depend only on $P$ and $R_{p}$, as incorporating information about individual stars would be too computationally expensive. In these cases, two planets with the same period and radius but host stars with vastly different properties would still be assigned the same completeness.

Furthermore, given that we are focused on specific period and radius bins, occurrence rates may be sensitive to the accuracy of a planet's membership in its correct bin. While orbital period is typically known to an accuracy of minutes or better, uncertainties in planet radius are significantly larger. First, measurement errors caused by fitting a transit model to a noisy light curve can cause the fitted ratio $R_{p}/R_{s}$ to differ from its true value. Second, and more significantly, uncertainty in the star's radius used to derive $R_{p}$ from $R_{p}/R_{s}$ directly leads to uncertainty in the planet's radius, even if $R_{p}/R_{s}$ is known exactly. As a result, a planet's ``observed'' radius bin may differ from its true radius bin, particularly if it is near the boundary between two bins. As in \citet{hsu19}, our forward model is able to take into account these measurement uncertainties by simulating both true and observed stellar and planetary radii, while most other studies assume that a planet's properties are known exactly.

\subsubsection{Step 1: Generate Planets}

We start by determining the number of planets to be simulated in our population. Given that the occurrence rate $f_{p,r}$ represents the average number of planets per star in period bin $p$ and radius bin $r$, and considering there are $N_{s}$ stars in the sample, the number of planets  in each bin can be drawn from a Poisson distribution with rate $\lambda = f_{p,r}N_{s}$.

Then, we assign each planet a star at random, and draw physical and orbital properties from model distributions. We draw the precise orbital period ($P$) and radius ($R_{p}$) uniformly in log period and log radius, constrained to be within the assigned bin. We assume circular orbits ($e = 0$), and assume the orbital inclinations ($i$) are uniformly distributed across the sky, drawing from $\cos{i} \sim U(0,1)$. 

We note that we do not take into account correlations in planet properties in multiplanet systems, and only assign a star to each planet for the purpose of attaining a stellar radius, mass, and other relevant parameters. In other words, planets are drawn completely independently of one another. Our assumption of circular orbits, while consistent with previous works, is also simplistic, and systems with a single transiting planet have been shown to have a different eccentricity distribution than systems with multiple planets \cite[e.g. a mean of $e \approx 0.3$ compared to 0.04;][]{xie16}. However, these choices are primarily due to the computational expensiveness of running the ABC forward model, restricting us to fit only a select number of bins at a time and thus preventing us from simulating full system architectures. \citet{bur15} also showed that incorporating nonzero eccentricity (assuming all planets have $e = 0.4$) had only a modest impact on occurrence rates, comparable to statistical errors.

\subsubsection{Step 2: Calculate Selection Effects}

\vspace{5pt}
\centerline{\textit{Transit Probability}} 
\vspace{5pt}

Many planets will be undetected simply because they do not cross our line of sight to the star. We use the planet's semi-major axis $a = (GM_{s}P^{2}/4\pi^{2})^{1/3}$ and inclination $i$ drawn previously to determine the planet's impact parameter

\begin{equation}
    b = \frac{a\cos{i}}{R_{s}},
\end{equation}

\noindent requiring that $b \leq 1$. In other words, the planet transits if the centre of the planet passes inside the disk of the star. As in \citet{hsu19}, we ignore the small number of transiting planets with $b > 1$, as large impact parameters are often associated with grazing eclipsing binaries and these planets are likely to be flagged as FPs. Thus, we set

\begin{equation}
    \eta_{\text{tr}} = 
    \begin{cases}
    1 & b \leq 1 \\
    0 & \text{otherwise.} \\
    \end{cases}
\end{equation}

\vspace{5pt}
\centerline{\textit{Recovery Efficiency}} 
\vspace{5pt}

We estimate the recoverability of each planet by taking into account pipeline search completeness, vetting completeness, and the probability that at least three transits occur in the \textit{Kepler} window. 

For search and vetting completeness, we use the combined search and vetting model as outlined in \S\ref{sec:comp}. We follow the same process outlined in \S\ref{sec:comp} to estimate each simulated planet's transit S/N and $N_{\text{tr}}$ to determine the corresponding $P_{\text{det}}$.

For the window probability, we use the binomial probability function described in \citet{bur15}

\begin{equation}
    \begin{split}
        P_{\text{win},\geq 3} = 1 & - (1 - f_{\text{duty}})^{M} - Mf_{\text{duty}}(1-f_{\text{duty}})^{M-1} \\
        & - \frac{M(M-1)}{2}f_{\text{duty}}^{2}(1-f_{\text{duty}})^{M-2}
    \end{split}
\end{equation}

\noindent where $M = T_{\text{obs}}/P$. Thus, we find the total recovery efficiency for a given planet as

\begin{equation}
    \eta_{\text{rec}} = P_{\text{det}} P_{\text{win},\geq 3}.
\end{equation}

\subsubsection{Step 3: Simulate Detected Exoplanet Population}

We determine if a planet is detected by drawing from a Bernoulli distribution with probability

\begin{equation}\label{eqn:net}
\eta_{\text{tot}} = \eta_{\text{tr}}\eta_{\text{rec}}.
\end{equation}

\noindent At this point, we remove all planets flagged as undetected from the simulation and focus the remainder of our analysis on the recovered population.

\subsubsection{Step 4: Incorporate Planet Radius Uncertainty}

We cannot assume that once a planet is detected, we also recover its true radius exactly. We take into account measurement errors caused by fitting a transit model as well as uncertainty in a host star's radius as in \citet{hsu19}.

First, we compute a planet's planet-to-star radius ratio $k = R_{p}/R_{s}$ using the true planet and stellar radius. Then, we draw an observed stellar radius $R_{s,\text{obs}}$ from two half-normal distributions, with median equal to the \citet{ber18} radius and widths equal to the upper and lower radius uncertainties. We also draw an observed $k_{\text{obs}}$ centred on the true $k$ based on the transit's S/N and the diagonal noise model of \citet{pri14}. Finally, we compute the observed planet radius as $R_{p,\text{obs}} = k_{\text{obs}} R_{s,\text{obs}}$.

We place each simulated planet into a new radius bin depending on the results of this process. In doing so, we create our final simulated exoplanet catalogue, to be compared with the actual catalogue produced from our \textit{Kepler} search.

\subsubsection{Step 5: Compare to Observed Population}

We generate summary statistics for each bin in both observed and simulated catalogues, 

\begin{equation}
    s_{k} = \frac{N_{k}}{N_{s}},
\end{equation}

\noindent where $N_{k}$ is the number of planets in the $k$th bin. We use the fraction of planets per star rather than the absolute number of planets so as to allow for differences in the choice of $N_{s}$ between catalogues. For instance, we could choose to run a quick inference by comparing our search results from all 96,280 FGK stars with a catalogue that simulates planets around only 10,000 FGK stars.

It is at this point that we apply our distance function to quantify the distance between summary statistics and thus assess the agreement between the simulated and observed planet catalogues.

\subsection{Distance Function}

When modeling only a single period-radius bin at a time, such as in \citet{hsu18}, the summary statistic for each catalogue is scalar. The choice of distance function may be simply

\begin{equation}
    \rho(s_{\text{obs}}, s_{\text{sim}}) = (s_{\text{obs}} - s_{\text{sim}})^{2},
\end{equation}

\noindent where $s = N/N_{s}$ is calculated for only the single bin of interest, and obs and sim refer to the observed and simulated catalogues respectively. However, when fitting multiple bins simultaneously, we use the distance suggestion of \citet{hsu19}, 

\begin{equation}
    \rho(s_{\text{obs},k},s_{\text{sim},k}) = \sum_{k}{\frac{|s_{\text{obs},k} - s_{\text{sim},k}|}{\sqrt{s_{\text{obs},k} + s_{\text{sim},k}}}},
\end{equation}

\noindent inspired by the Canberra distance \citep{lan67}. \citet{hsu19} found that this distance allowed ABC to converge more rapidly than other tested functions. This function also weights the absolute value of the differences in $s_{k}$ by the square root of the sum, resulting in a similar fractional error in occurrence rates for all bins rather than a similar absolute error.

\subsection{Model Verification}\label{sec:test}

With our ABC framework set, we justified the number of bins to fit at once and verified that the algorithm was able to recover occurrence rates accurately with the appropriate choices.

\begin{figure*}[t!]
\centering
\includegraphics[width=0.8\textwidth]{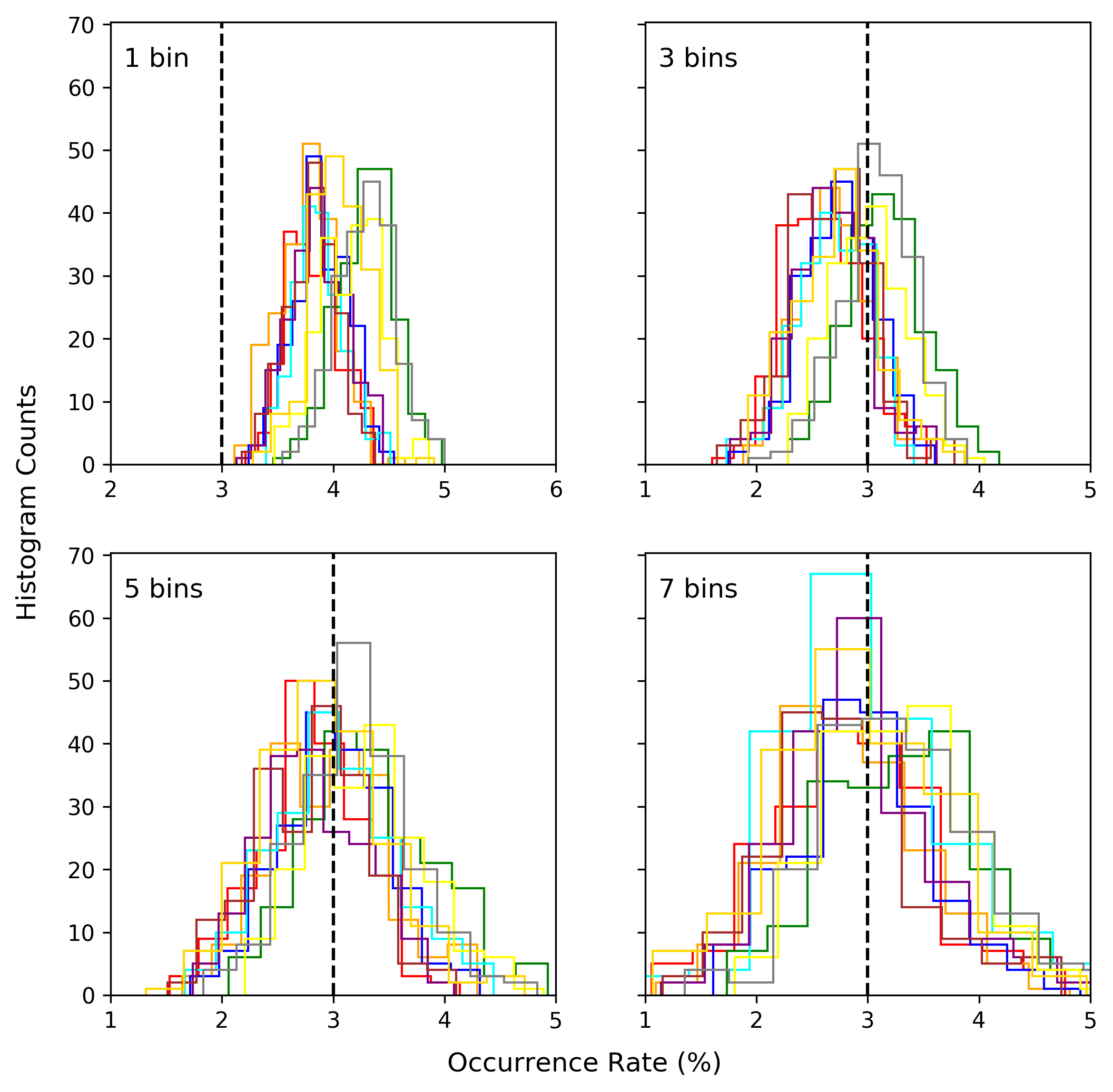}
\caption{Results from testing how the recovery of the $6.25 - 12.5$ day, $1 - 1.41$ $R_{\bigoplus}$ simulated occurrence rate ($f = 0.03$, or 3$\%$) changes depending on the number of bins fit simultaneously. Each colour corresponds to 1 of 10 simulated planet catalogues.}\label{fig:test}
\end{figure*}

Had we not incorporated planet radius uncertainty into the forward model, the placement of each simulated planet into a specific period-radius bin would be without ambiguity. The occurrence rates of each bin would not affect those of others, and thus fitting only one bin at a time would be an obvious choice due to computational efficiency.

Because our simulator takes into account measurement error, it may place a planet into a radius bin different from its true bin. If two neighbouring bins have different occurrence rates, the number of planets exchanged across the radius boundary may be asymmetric. Furthermore, the edge bins being fit will display noticeable bias. As the simulator does not simulate planets with true radii above the upper limit of the top bin and below the lower limit of the bottom bin, the exchange of planets over these radii limits will be strictly one sided, and their occurrence rates will be over estimated.

These considerations necessitate the fitting of multiple bins simultaneously. However, fitting more parameters comes at the cost of the performance of the ABC-PMC algorithm, as it becomes less likely that the proposed values for all parameters will result in good agreement between the observed and simulated catalogues. Both the width of the ABC posterior and the computational time required for the algorithm to achieve convergence will increase significantly.

To explore these issues, we used our forward model to simulate 10 ``true'' planet catalogues in the $6.25 - 12.5$ day period range using the full 96,280 star sample. We used occurrence rates of $f = \{0.06, 0.05, 0.04, 0.03, 0.02, 0.01, 0.005\}$ for bins with boundaries $R_{p} = \{0.35, 0.5, 0.71, 1, 1.41,$ $2, 2.83, 4\}$ $R_{\bigoplus}$. We fit a total of one, three, five, and seven bins simultaneously, centred on the $1.25-1.5R_{\bigoplus}$ bin ($f = 0.03$, or $3\%$), for each simulated catalogue. As inputs to \texttt{cosmoabc}, we set the number of particles for the initial generation at 500 and all subsequent generations at 200. We considered the system converged when at least 2000 draws (10 times the size of each generation) were required to construct the next generation.

Fig. \ref{fig:test} shows the final ABC posterior for the $6.25-12.5$ day, $1-1.41 R_{\bigoplus}$ occurrence rate after each run. As expected, the one-bin fit consistently overestimates the true occurrence rate due to the fact that simulated planets can only leak out of the bin. The average absolute difference between the ABC posterior median and the true 3$\%$ occurrence rate was 0.98$\%$. The three-, five-, and seven-bin fits all show significant improvement, with ABC posteriors well clustered around the true occurrence rate. Average absolute differences were $0.24\%$, $0.15\%$, and $0.17\%$ respectively. We also observe the expected widening of the ABC posterior with more bins.

When examining the results for the edge bins in each multibin-fit run, we confirmed that they tended to be overestimated compared to the interior bins. This was especially apparent for the bottom-edge bins, likely due to the fact that they were assigned the highest occurrence rates and thus had more outward leakage of planets than inward. These considerations prompted us to exclude the results for the two edge bins when performing multibin fits, and only report the results for the interior bins. 

Overall, we agree with the conclusions of \citet{hsu19} that five to seven radius bins are the optimal choice, and that one should be careful when considering the results of edge bins. 

\section{Occurrence Rate Results}\label{sec:results}

Our baseline exoplanet occurrence rates are defined using the combined FGK sample without reliability, as well as F, G, and K stars separately, using a period-radius grid with logarithmically spaced bin edges of $P = \{0.78, 1.63, 3.13, 6.25, 12.5, 25, 50, 100, 200, 400\}$ days and $R_{p} = \{0.5, 0.71, 1, 1.41, 2, 2.83, 4, 5.66, 8, 11.31, \linebreak 16\} R_{\bigoplus}$. For our FGK sample, which is expected to be the best constrained due to having the largest number of bins populated with planets, we produce additional occurrence rates after taking into account the reliability of our pipeline.

We used our investigations of the multibin fits to determine the final setup for our full occurrence rate estimates. Because our interest is in planets with radii down to 0.5 $R_{\bigoplus}$, our final results involve fitting additional 0.35 - 0.5$R_{\bigoplus}$ bins for the sole purpose of acting as an edge bin to ensure accuracy for the 0.5 - 0.71$R_{\bigoplus}$ bin. We are also interested in planets with radii up to 16 $R_{\bigoplus}$, but given that planets with $R_{p} > 16$ $R_{\bigoplus}$ are rare, we do not expect the same bias to be present and keep our results for the 11.31 - 16 $R_{\bigoplus}$ bins as is. Therefore, we report our results using five-bin fits with radius boundaries $R_{p} = \{0.35,0.5,0.71,1,1.41,2\}$, $\{1,1.41,2,2.83,4,5.66\}$, and $\{2.83,4,5.66,8,11.31,16\}$ $R_{\bigoplus}$ for each period range. 5-bin fits were chosen as a balance between minimizing edge-bin bias while avoiding the unnecessary broadening of the ABC posterior. After removing the edge bins (with the exception of the 11.31 - 16 $R_{\bigoplus}$ bin) from each subset, the entire $0.5 < R_{p} < 16$ $R_{\bigoplus}$ radius range of interest is covered.

\begin{figure*}
\centering
\includegraphics[width=0.6\textwidth]{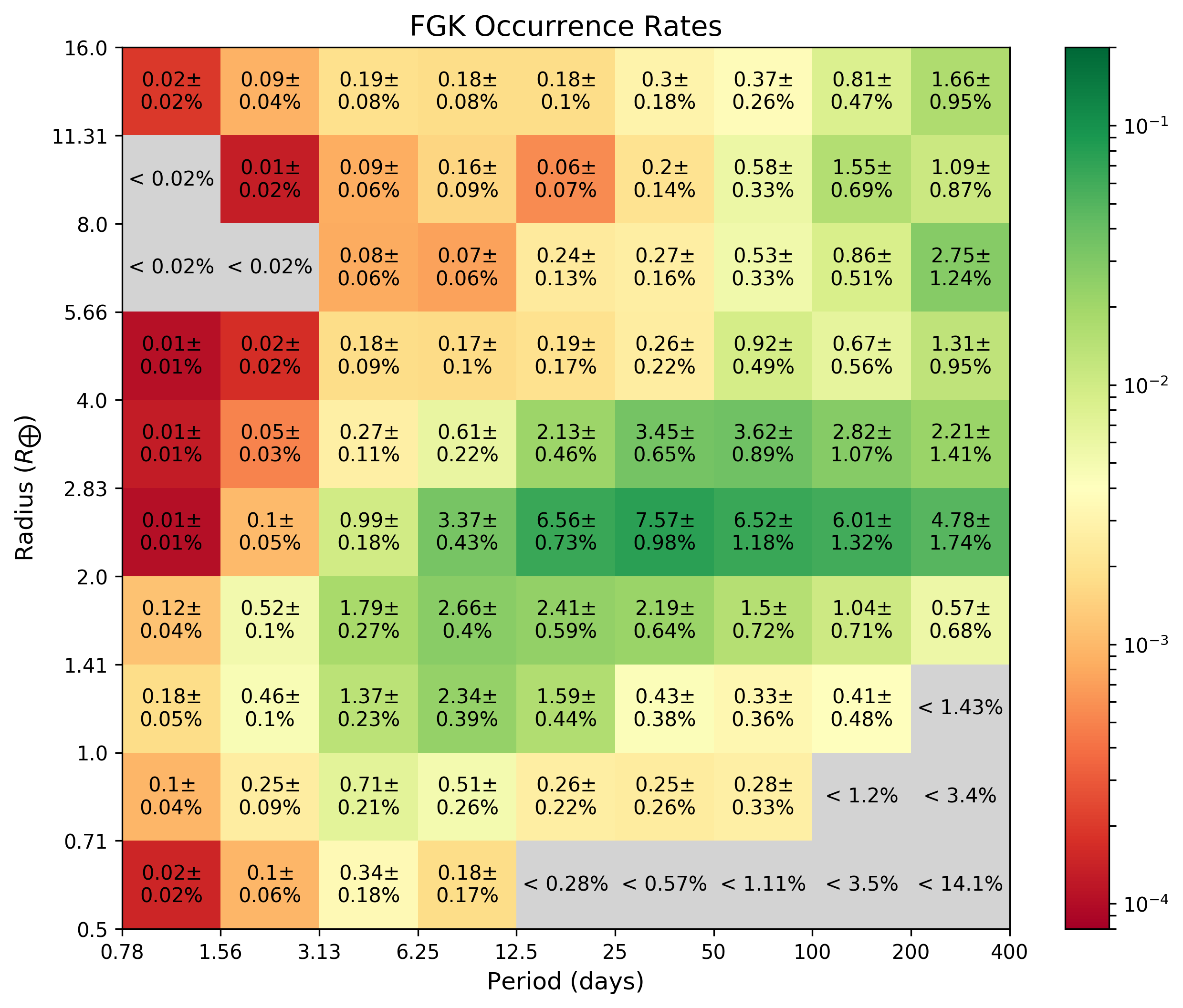}
\caption{Occurrence rate estimates for FGK-type stars. The number of planets per star is given in percentage ($10^{-2}$) and is the median of the ABC posterior. Uncertainties are the larger of the lower and upper uncertainties, calculated as as the difference between the median and 15.9th and 84.1th percentiles, respectively. Bins with no detected planets are in grey, with only the upper limit (84.1th percentile) shown.}\label{fig:FGKrates}
\end{figure*}

\begin{figure*}
\centering
\includegraphics[width=0.6\textwidth]{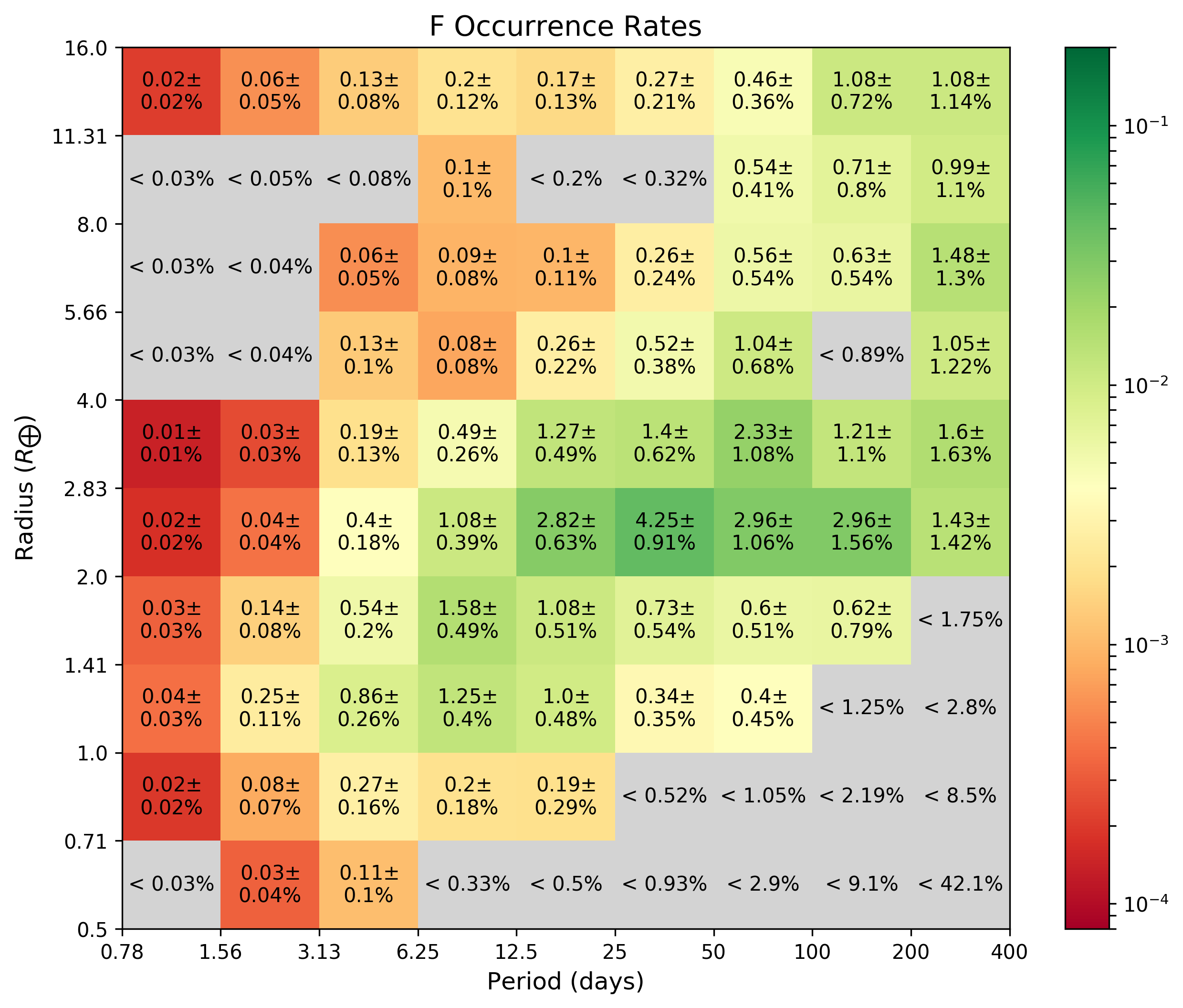}
\caption{Same as Fig. \ref{fig:FGKrates}, but for F-type stars only.}\label{fig:Frates}
\end{figure*}

\begin{figure*}
\centering
\includegraphics[width=0.6\textwidth]{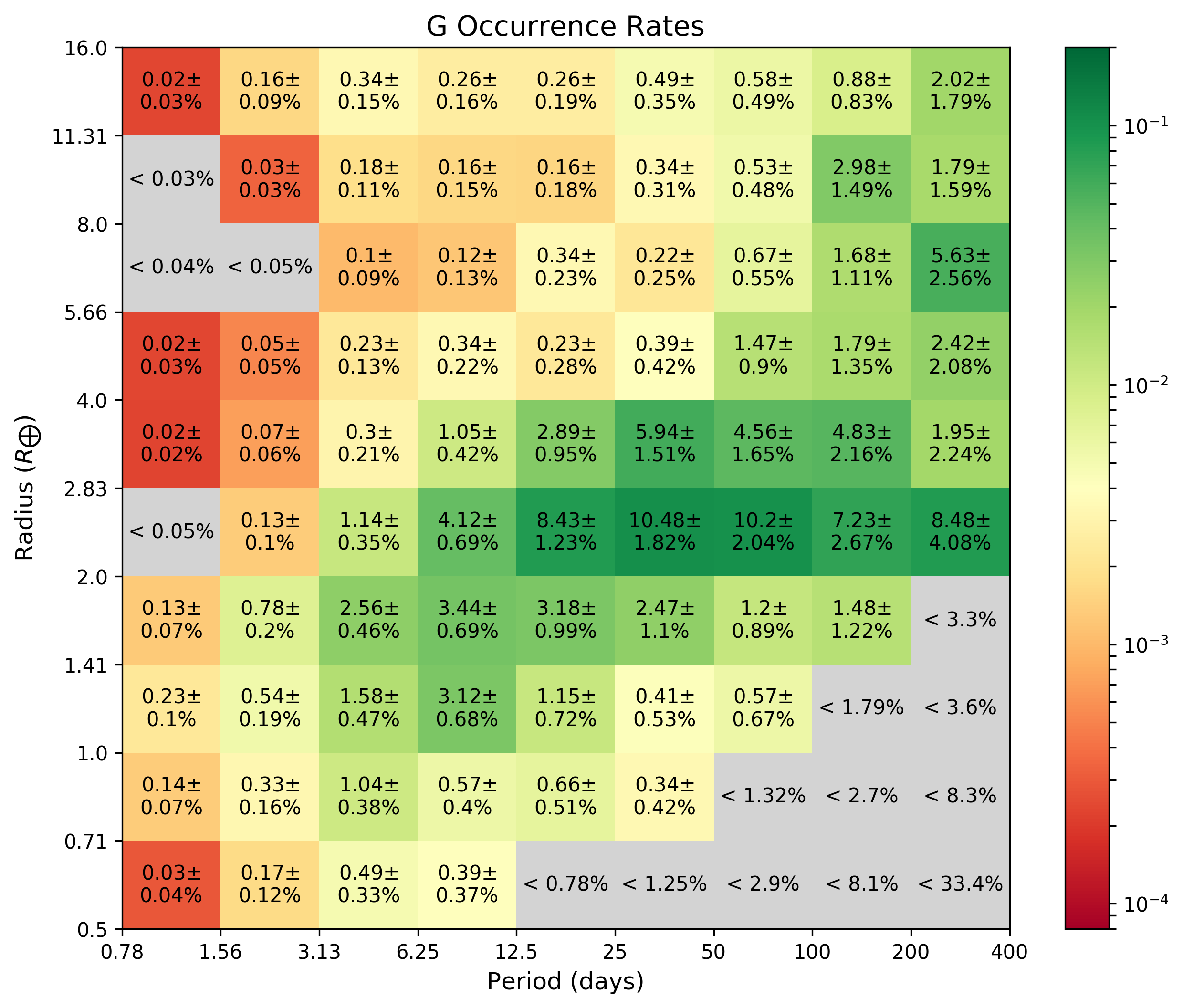}
\caption{Same as Fig. \ref{fig:FGKrates}, but for G-type stars only.}\label{fig:Grates}
\end{figure*}

\begin{figure*}
\centering
\includegraphics[width=0.6\textwidth]{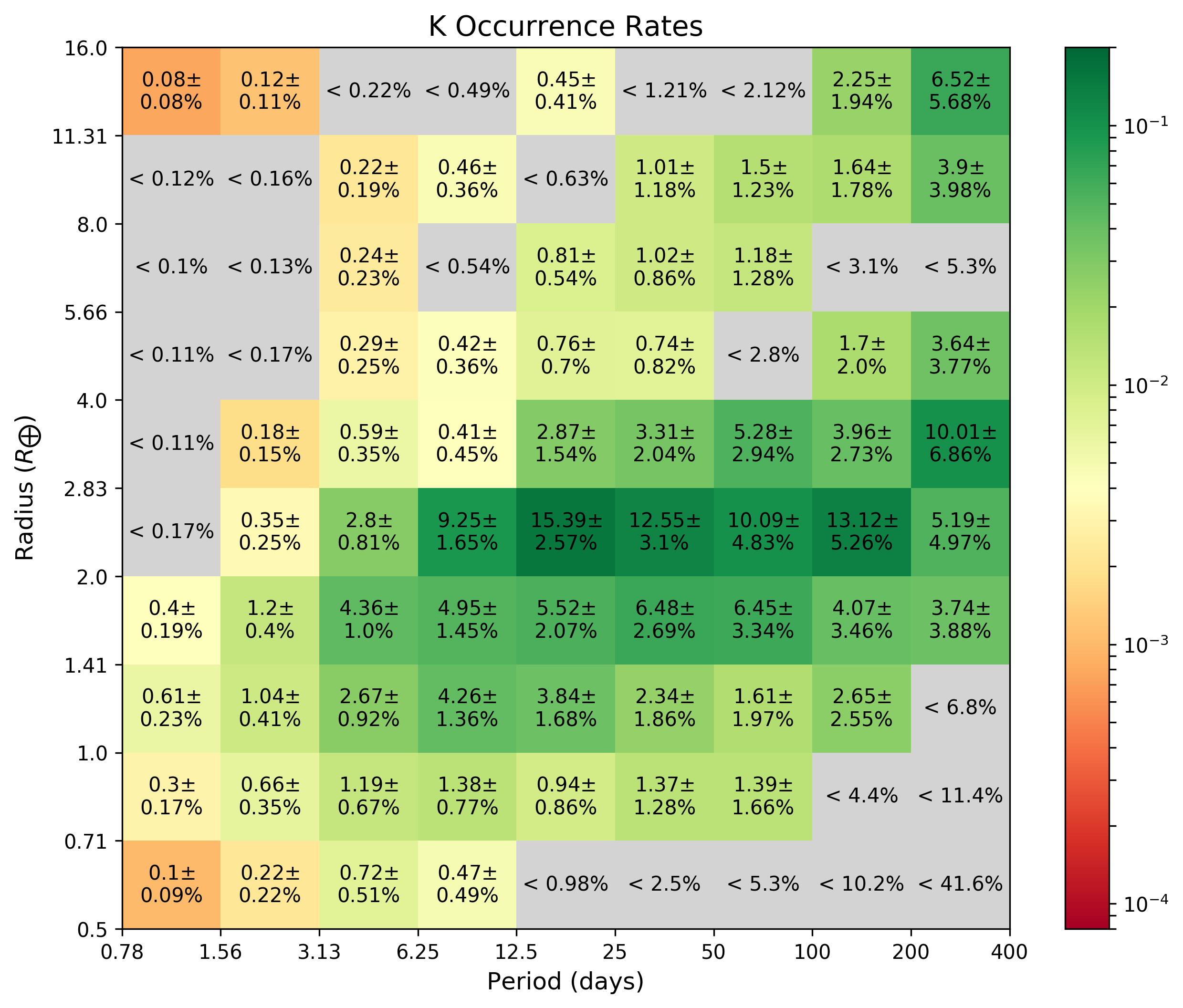}
\caption{Same as Fig. \ref{fig:FGKrates}, but for K-type stars only.}\label{fig:Krates}
\end{figure*}

Our final FGK, F, G, and K results are given in Table \ref{tbl:FGKrates}. We report the occurrence rate as the median of the ABC posterior for each $f_{p,r}$, with the difference between the median and 15.9th and 84.1th percentiles as the lower and upper uncertainties, respectively. For bins with zero detected planets, we report only the upper limit (84.1th percentile). We plot baseline FGK occurrence rates in Fig. \ref{fig:FGKrates}, followed by occurrence rates for F-, G-, and K-type stars in Figs. \ref{fig:Frates}, \ref{fig:Grates}, and \ref{fig:Krates} respectively. Uncertainties are represented by the larger of the lower and upper uncertainties for each bin. We set the colour scale to be the same in all four plots for more direct visual comparison.

We also follow the recommendations of the Study Analysis Group (SAG) 13 of the NASA Exoplanet Exploration Program Analysis Group (ExoPAG)\footnote{https://exoplanets.nasa.gov/system/presentations/files/67\textunderscore \linebreak Belikov\textunderscore SAG13\textunderscore ExoPAG16\textunderscore draft\textunderscore v4.pdf} and estimate F, G, and K occurrence rates on a grid with bin edges of $P = \{10, 20, 40, 80, 160, 320, 640\}$ days and $R_{p} = \{0.67, 1, 1.5, 2.25, 3.38, 5.06, 7.59, 11.39, 17.09\} R_{\bigoplus}$. These results are presented in Table \ref{tbl:exopag}.

\subsection{General Comparison to Previous Works}

We compare our occurrence rates with those reported by other studies, choosing \citet{ful17}, \citet{mul15a}, \citet{pet13}, and \citet{fre13} as targets for comparison on the basis of having the most similar period and radius ranges. Given that occurrence rate studies calculate occurrence rates with different methods, account for completeness in different ways, use catalogues based on different amounts of available \textit{Kepler} photometry, and more, direct comparison is difficult. Additionally, \citet{pet13} inferred occurrence rates using only the first planet found in each system. Consequently, their occurrence rates are not estimates of the average number of planets per star as in the other works. Recognizing these challenges, \citet{ful17} compared the ratios of occurrence rates between bins rather than the absolute occurrence rates of individual bins. We adopt the same approach here.

Table \ref{tbl:comparisonFGK} shows our FGK results compared to \citet{fre13} (FGK), \citet{mul15a} (FGKM), and \citet{ful17} (FGK) while Table \ref{tbl:comparisonG} shows our G results compared to \citet{pet13} (GK) and \citet{mul15a} (G). We find that the main discrepancy is that the ratios of occurrence rates for $2-4$ $R_{\bigoplus}$ planets to $1-2$ $R_{\bigoplus}$ planets, across all period ranges and for both FGK- and G-type stars, are higher than all previous works. This is most noticeable when comparing to the older studies of \citet{pet13} and \citet{fre13}. \citet{pet13} found that $1-2$ $R_{\bigoplus}$ planets within 50 days are 1.2 times more common than those with radii $2-4$ $R_{\bigoplus}$. \citet{fre13} found a similar ratio of 1.1. Meanwhile, \citet{mul15a} found that planets with $1-2$ $R_{\bigoplus}$ are less common, with ratios of 0.9 (FGKM stars) and 0.6 (GK). Our results are even more favoured toward the larger radius bin, with ratios of 0.6 (FGK) and 0.4 (G).

\begin{table*}
    \centering
    \begin{threeparttable}[b]
    \caption{Rough comparisons for FGK-type stars. Our results are FGK occurrence rates marginalized over periods down to 0.78125 days. F17 results are taken from Table 5 in \citet{ful17}. M15 results are taken from Table 6 in \citet{mul15a}, summing bins down to 0.68 days. F13 results are taken from Table 3 in \citet{fre13}, reported down to 0.8 days. For all results that involved summing multiple bins, we used the propagation of error to estimate uncertainty.}
    \label{tbl:comparisonFGK}
    \begin{tabular}{c|c|c|c|c|c}
    \hline
    \hline
    $R_{p}$ ($R_{\bigoplus}$) & $P$ (days) & \textbf{This work} (\%) & F17 (\%)& M15 (\%)& F13 (\%)\\
    \hline
    $1-2$ & $< 50$ & $16.2_{-1.1}^{+1.2}$ & - & $16.3\pm0.7$ & $19.4\pm2.0$\tnote{1} \\
    $2-2.8$ & $< 50$ & $18.6_{-1.2}^{+1.3}$ & $19.4\pm1.4$ & $12.7\pm0.5$ & -\\
    $2-4$ & $< 50$ & $25.2_{-1.5}^{+1.5}$ & $25.4\pm1.6$ & $18.6\pm0.6$ & $18.3\pm1.3$ \\
    $4-8$ & $< 50$ & $1.6_{-0.3}^{+0.4}$ & - & $3.1\pm0.2$ & - \\
    $8-16$ & $< 50$ & $1.6_{-0.3}^{+0.3}$ & - & $2.0\pm0.2$ & - \\
    \hline
    $1-2$ & $< 100$ & $18.1_{-1.4}^{+1.4}$ & - & $16.3\pm0.7$\tnote{2} & $23.0\pm2.4$\tnote{1,2} \\
    $1.4-2.8$ & $< 100$ & $36.5_{-2.0}^{+2.1}$ & $43.1\pm2.2$ & $26.7\pm0.8$\tnote{2} & - \\
    $2-4$ & $< 100$ & $35.4_{-2.1}^{+2.1}$ & $36.6\pm2.2$ & $23.0\pm0.8$\tnote{2} & $23.5\pm1.6$\tnote{2} \\
    $4-8$ & $< 100$ & $3.1_{-0.6}^{+6.7}$ & - & $4.4\pm0.3$\tnote{2} & - \\
    $8-16$ & $<100$ & $2.6_{-0.4}^{+0.5}$ & - & $2.6\pm0.2$\tnote{2} & -\\
    \end{tabular}
    \begin{tablenotes}
    \item[1] $1.25 - 2$ $R_{\bigoplus}$
    \item[2] $P < 85$ days
    \end{tablenotes}
    \end{threeparttable}
\end{table*}

    \begin{table*}
    \centering
    \begin{threeparttable}[b]
    \caption{Rough comparisons for G-type stars. Our results are G occurrence rates marginalized over periods down to 6.25 days (lower bound chosen to match the lower bound of \citet{pet13} results). M15 results are taken from Table 7 in \citet{mul15a}, summing bins down to 5.8 days. P13 results are taken from Fig. 2 in \citet{pet13}, summing bins down to 6.25 days. For all results that involved summing multiple bins, we used the propagation of error to estimate uncertainty.}
    \label{tbl:comparisonG}
    \begin{tabular}{c|c|c|c|c}
    \hline
    \hline
        $R_{p}$ ($R_{\bigoplus}$) & $P$ (days) & \textbf{This work} (\%)& M15 (\%) & P13\tnote{1} (\%) \\
        \hline
        $1-2$ & $< 50$ & $14.1_{-1.9}^{+2.0}$ & $12.9\pm0.9$ & $19.2\pm1.7$ \\
        $2-4$ & $< 50$ & $33.2_{-3.0}^{+3.0}$ & $20.8\pm1.0$ & $16.4\pm1.2$  \\
        $4-8$ & $< 50$ & $1.8_{-0.5}^{+0.6}$ & $3.3\pm0.4$ & $1.5\pm0.3$ \\
        $8-16$ & $< 50$ & $1.8_{-0.5}^{+0.5}$ & $1.8\pm0.3$ & $1.0\pm0.4$\\
        \hline
        $1-2$ & $< 100$ & $16.0_{-2.0}^{+2.2}$ & $16.0\pm1.4$\tnote{2} & $25.0\pm2.3$ \\
        $2-4$ & $< 100$ & $48.0_{-3.8}^{+4.0}$ & $25.2\pm1.2$\tnote{2} & $24.1\pm1.8$  \\
        $4-8$ & $< 100$ & $4.1_{-1.0}^{+1.2}$ & $4.8\pm0.6$\tnote{2} & $2.8\pm0.7$ \\
        $8-16$ & $<100$ & $3.1_{-1.8}^{+2.3}$ & $2.5\pm0.4$\tnote{2} & $1.6\pm0.5$ \\
     \end{tabular}
    \begin{tablenotes}
    \item[1] Fraction of stars with planets instead of number of planets per star
    \item[2] $P < 85$ days
    \end{tablenotes}
    \end{threeparttable}
\end{table*}

\citet{ful17} also found a lower fraction of planets below $2$ $R_{\bigoplus}$ than older works. In particular, they found a $P < 100$ day, $1.41 - 2$ $R_{\bigoplus}$/$2 - 2.83$ $R_{\bigoplus}$ ratio of 0.6, whereas \citet{pet13} found 1.3. They explained this difference using their knowledge of a gap in the radius distribution between 1.5 and 2 $R_{\bigoplus}$ and a peak near $\sim$2.5 $R_{\bigoplus}$, which were both revealed in their study. Because these features were recovered in part due to their use of more precise stellar radii from spectroscopy, they suggested that the large ($\approx40$\%) radius uncertainties from photometry alone would scatter planets with true sizes between 2 and 2.83 $R_{\bigoplus}$ to the $1.41-2$ $R_{\bigoplus}$ bin, both filling the gap and reducing the peak. Given that we used updated stellar radii from the \citet{ber18} catalogue, which brought typical radius uncertainties down to $\approx8$\% and found a similarly low ratio (0.4), the same explanation likely applies here. Thus, in combination with the results of \citet{ful17}, our results emphasize the sensitivity of planet occurrence rates to accurate stellar radii. We also argue that our results are more robust than previous works, given our use of updated stellar radii in combination with direct incorporation of planet radius uncertainties into our occurrence rates.

In the following sections, we discuss further comparisons to previous works in the context of interesting and informative features previously uncovered from exoplanet population analysis and occurrence rate estimates.

\clearpage

\subsection{Dependence on Stellar Effective Temperature}\label{sec:temperature}

Marginalizing over the entire period-radius grid, we find occurrence rates of $0.89_{-0.16}^{+0.23}$ planets per F-type star, $1.67_{-0.16}^{+0.21}$ planets per G-type star, and $2.56_{-0.24}^{+0.29}$ per K-type star, compared to $1.06_{-0.07}^{+0.09}$ for the combined FGK sample.\footnote{The FGK rate is slightly lower than what would be found if one were to combine each F, G, and K occurrence rate (after weighting by stellar sample size). This is because the larger sample size of the FGK catalogue produces more constrained occurrence rates, especially in the low-completeness and low-planet-detection regime where only upper limits can be reported.} For planets within 200 days (i.e. omitting the $200 - 400$ day bins which have low completeness and little to no planet detections below 2 $R_{\bigoplus}$), we find occurrence rates of $0.53_{-0.05}^{+0.06}$ (F), $1.17_{-0.07}^{+0.08}$ (G), $1.84_{-0.13}^{+0.15}$ (K), and $0.81_{-0.04}^{+0.04}$ (FGK). Our results indicate a statistically significant trend of increasing planet occurrence toward cooler stars.

\begin{figure}[t!]
\centering
\includegraphics[width=\linewidth]{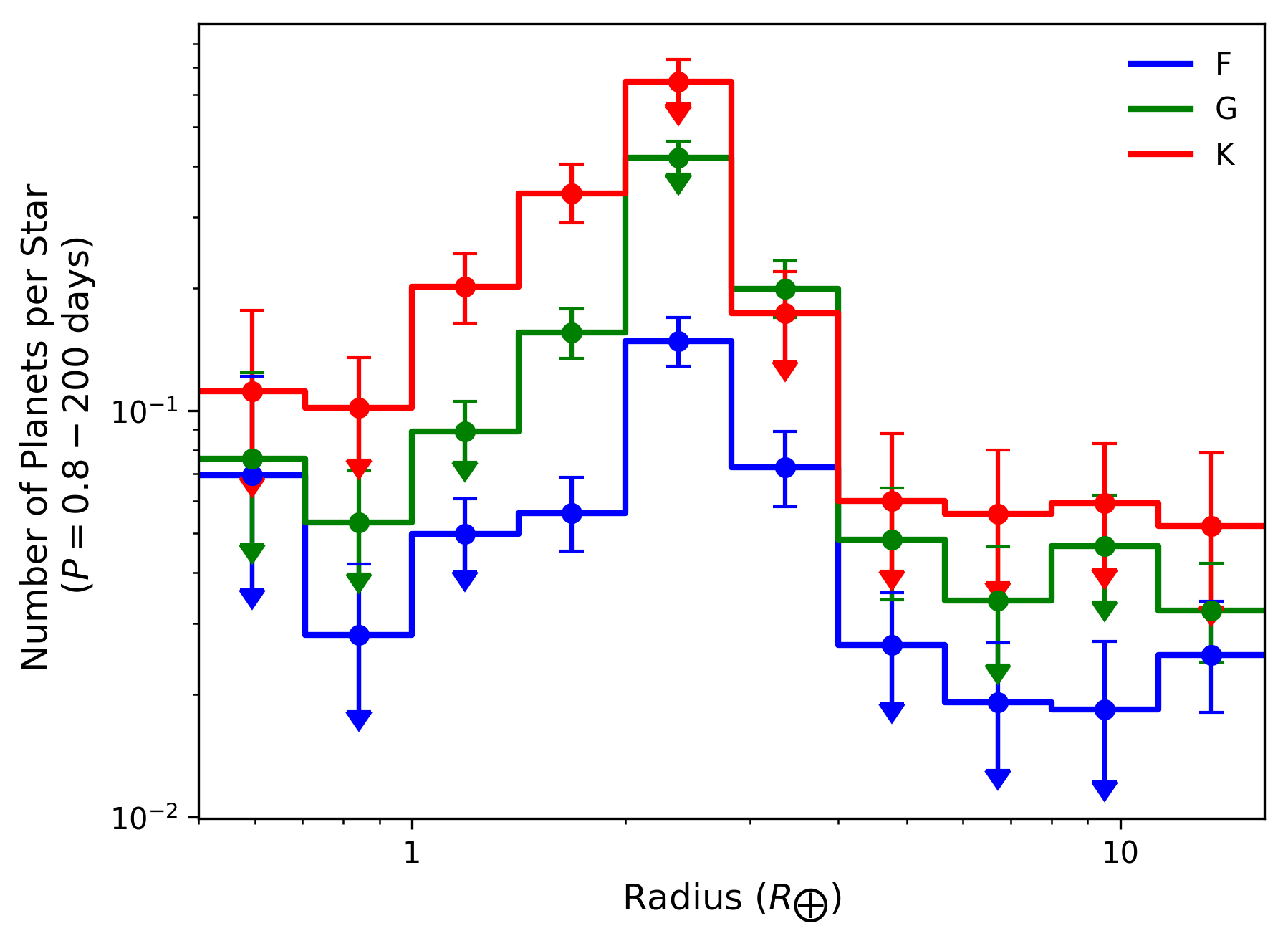}
\caption{Occurrence rates for planets within 200 days as a function of radius for F-, G-, and K-type stars}\label{fig:stellar}
\end{figure}

Fig. \ref{fig:stellar} shows the entire radius distribution marginalized over $P < 200$ days for each stellar type. We indicate any estimates that involved marginalizing over a bin with no planet detections with a downward pointing arrow, representing an upper limit. For small planets, our results are in strong agreement with \citet{how12} and \citet{mul15a} that occurrence rates increase substantially toward cooler stars. Overall, for $R_{p} = 1 - 2.83$ $R_{\bigoplus}$ and $P < 200$ days, we find occurrence rates of $0.26_{-0.02}^{+0.03}$ (F), $0.67_{-0.05}^{+0.05}$ (G), and $1.20_{-0.10}^{+0.11}$ (K). 

\noindent In other words, small planets around K-type stars are about twice as abundant than around G-type stars, and five times as abundant than around F-type stars. We also agree with \citet{mul15b} that each distribution shows a clear drop-off in planets beyond $2.83$ $R_{\bigoplus}$. Note that some measurements on both sides of the transition are only upper limits, but this is due to a lack of planets in the $0.78 -1.56$ day period range, which have upper limits of less than 0.002 planets per star. The existence of the drop-off is not dependent on their contribution.

For planets beyond $2.83$ $R_{\bigoplus}$, we find that the trend becomes less clear. The $2.83 - 4$ $R_{\bigoplus}$ bin demonstrates no statistically significant difference between G and K occurrence rates, with a G-K difference of $0.02_{-0.05}^{+0.05}$ planets per star yet a K-F difference of $0.10_{-0.04}^{+0.05}$. Over the same radius bin, \citet{mul15b} also found that G and K occurrence rates were indistinguishable while still significantly more common than around F type stars. At larger radii where the distributions flatten out for all stellar types, we do not attempt to interpret trends, as the majority of occurrence rate estimates involve summing bins without planet detections and thus only represent upper limits.

\subsection{Dependence on Planet Radius}\label{sec:radius}

\begin{figure}[t!]
\centering
\includegraphics[width=\linewidth]{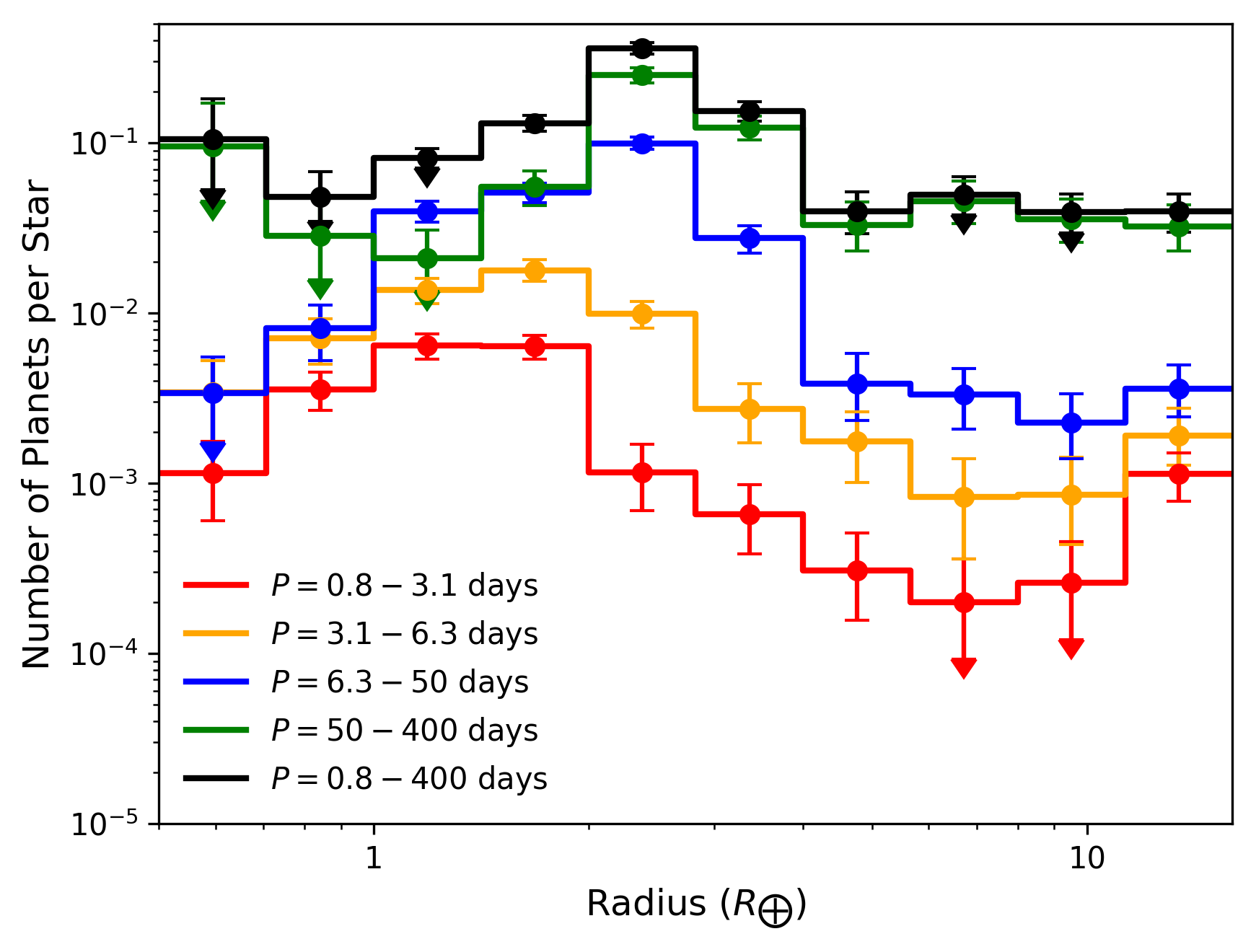}
\caption{FGK occurrence rates as a function of radius, marginalized over different period ranges.}\label{fig:radius}
\end{figure}

Our FGK occurrence rate results marginalized over different period ranges are shown in Fig. \ref{fig:radius}.

First, we note that we recover the ``Neptune desert,'' which is a dearth in Neptune- to sub-Jupiter-sized exoplanets in close-in orbits ($P \lesssim 3$ days) that has been noted and studied in many previous works \cite[e.g.][]{sza11, maz16, owe18}. On the lower-mass end of the desert, several studies have emphasized the role of photoevaporation on the mass loss of highly irradiated planets, while the dearth at the higher-mass end likely requires a different explanation \citep{ion18} such as tidal disruption that results from high-eccentricity migration \citep{owe18}. In particular, we observe a significant decrease in occurrence rates for planets between 4 and 11.31 $R_{\bigoplus}$ and the shortest orbital periods ($0.78 - 3.13$ days), but not larger planets, in line with expectations. Note that we did not detect any planets in the $5.66 - 11.31$ $R_{\bigoplus}$, $0.78 - 1.56$ day range, nor in the $5.66 - 8.0$ $R_{\bigoplus}$, $1.56 - 3.13$ day bins, so the dip is likely even lower than indicated in the plot. Meanwhile, at higher orbital periods (especially beyond 6.3 days), the distribution is flat over the same radii.

We now turn to smaller planets. An informative feature recently noted in exoplanet radius distributions is a gap between $\sim$1.5$ - 2$ $R_{\bigoplus}$ for planets within $P < 100$ days, also known as the ``radius valley'' \citep{ful17, van18}. The gap is accompanied by two peaks in radius, near $\sim$1.3 $R_{\bigoplus}$ (super-Earths) and $\sim$2.4 $R_{\bigoplus}$ (sub-Neptunes). This bimodal distribution had been predicted years earlier by numerical models involving the atmospheric erosion of highly irradiated low-mass planets \citep{lop13, owe13}, while its statistical significance was not established by completeness-corrected observations until \citet{ful17}, hereafter F17. F17 attributed the revelation of the radius gap to their use of precise stellar radius measurements from the California-Kepler Survey \cite[CKS][]{pet17}.

\begin{figure}[t!]
\centering
\includegraphics[width=\linewidth]{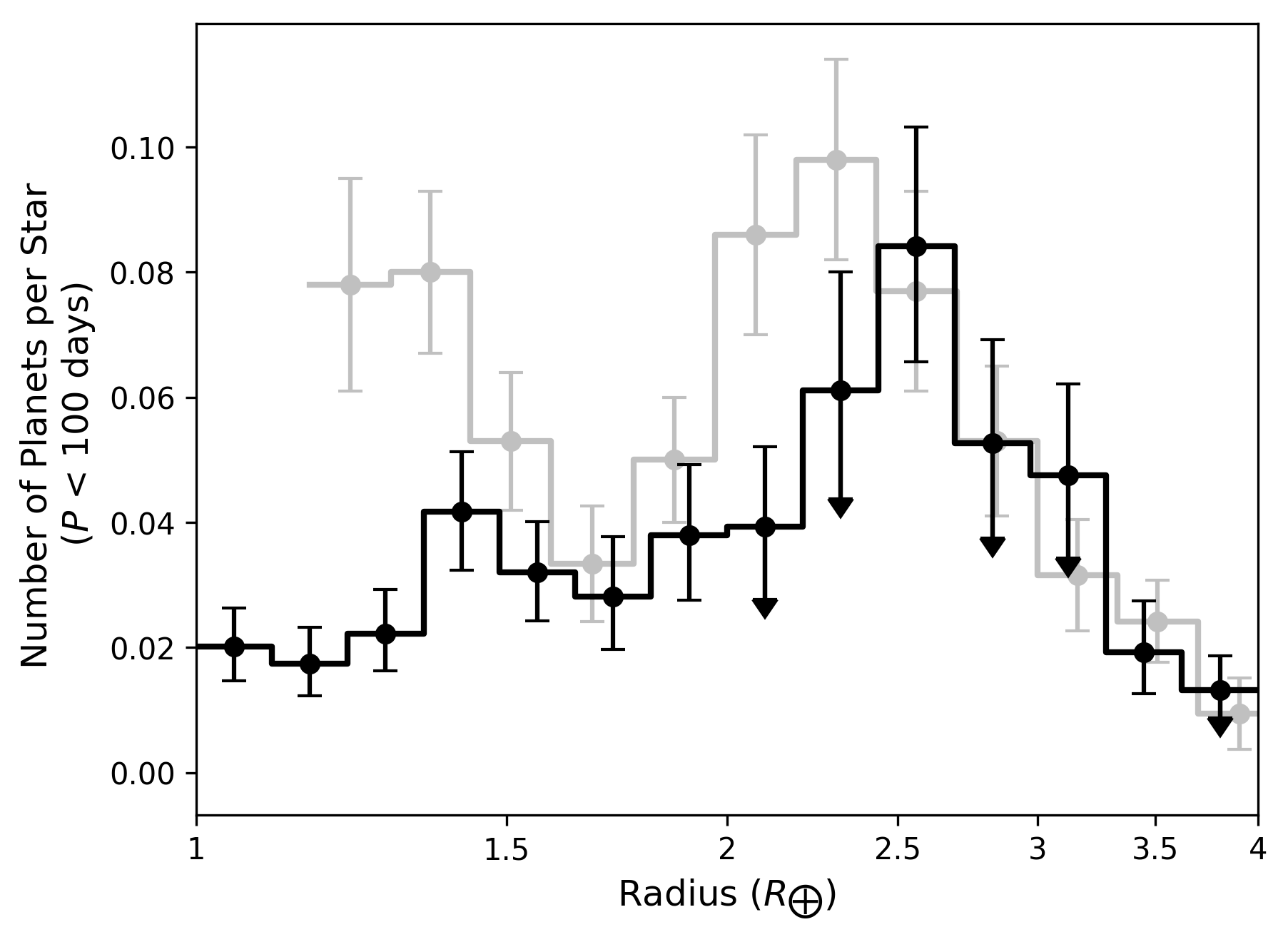}
\includegraphics[width=\linewidth]{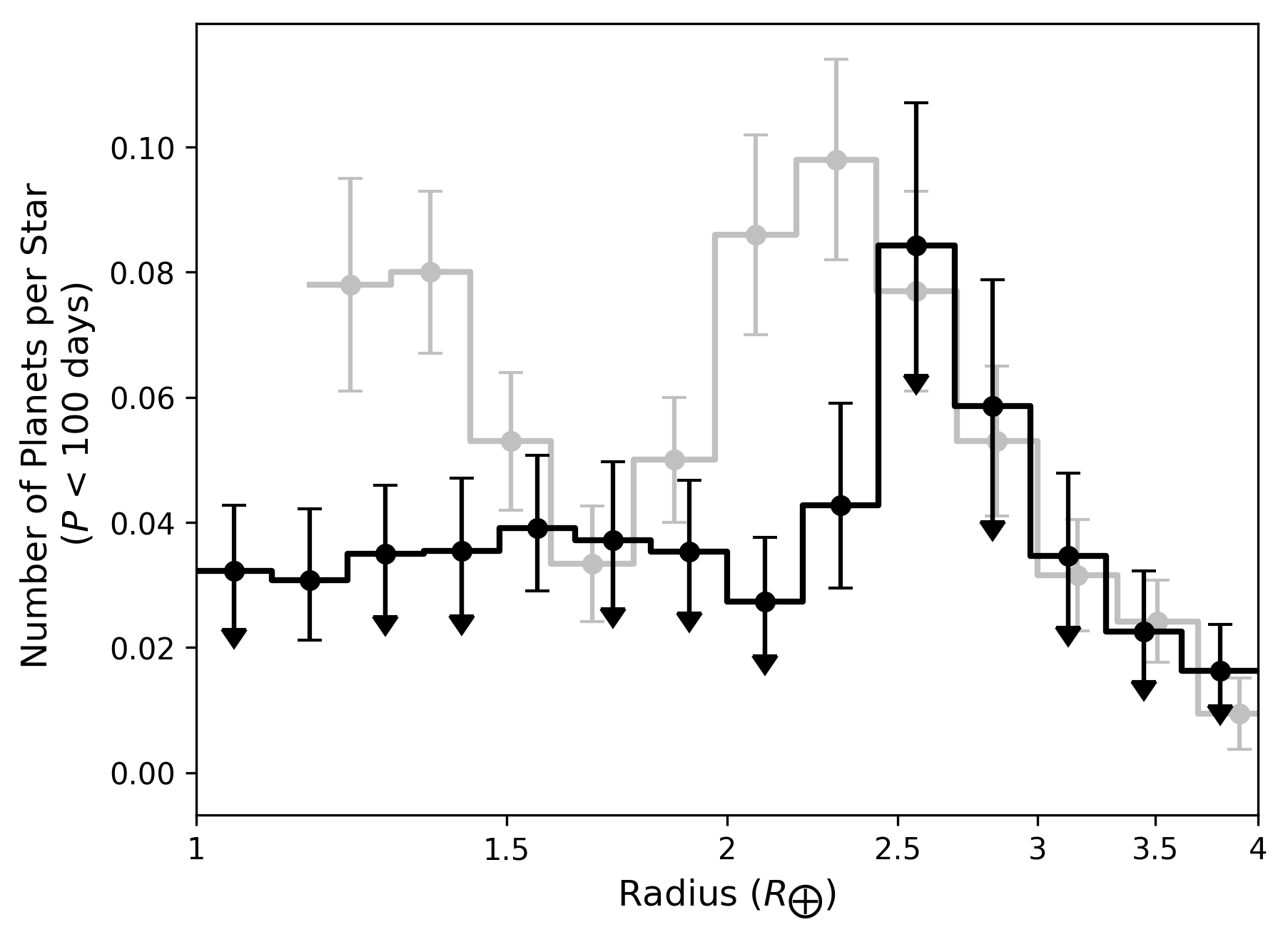}
\caption{Top: $P < 100$ day occurrence rates using radius bins smaller than our baseline study in order to compare to the \citet{ful17} radius valley. The occurrence rates from \citet{ful17} are shown in light grey down to 1.16 $R_{\bigoplus}$, beyond which results were not reported due to low completeness. Bottom: The same as above, but after applying a cut of $Kp < 14.2$ to our FGK sample.}\label{fig:valley}
\end{figure}

Our bin sizes were not small enough to be able to resolve this feature. This was a consequence of choosing logarithmically spaced bins large enough to be appropriate for the entire grid down to 0.5 $R_{\bigoplus}$ and out to 400 days. Thus, we recomputed $0.78 - 100$ day occurrence rates using the same $P = \{0.78, 1.56, 3.13, 6.25, 12.5, 25, 50, 100\}$ day period bins, but much finer bins in radius space over the regime of interest: $R_{p} = \{1, 1.10, 1.21, 1.35, 1.49, 1.64, 1.81, 2, 2.21, 2.44, 2.69, 2.97, \newline 3.28, 3.62, 4\}$ $R_{\bigoplus}$. The resulting distribution marginalized over the entire period range is compared with occurrence rates from Table 3 of F17, shown in the top panel of Fig. \ref{fig:valley}.

While we do find some evidence for a first peak near $\sim$1.4 $R_{\bigoplus}$, a minimum at $\sim$1.7 $R_{\bigoplus}$, and a second peak at $\sim$2.6 $R_{\bigoplus}$, differences between our distribution and that of F17 are obvious. Planets smaller than $1.5$ $R_{\bigoplus}$ are considerably less abundant according to our sample, and we cannot confirm that our radius valley is statistically significant from these results alone. Meanwhile, our second peak is shifted to a higher radius than in F17, and it is twice as tall as the first peak rather than being comparable in height.

It should be noted that F17 only considered planets with hosts fainter than $Kp = 14.2$, given that the core sample of the CKS is magnitude limited, and the distribution of CKS planet radii above and below $Kp = 14.2$ is statistically different. In order to see if this could explain the differences between our results, we recalculated our occurrence rates using only our FGK stars with $Kp < 14.2$, which reduced our stellar sample from 96,280 to 30,688, and our $P < 100$ day planet population from 2377 to 833. This is shown in the bottom panel of Fig. \ref{fig:valley}. While interpreting the results is difficult given that the occurrence rates are less well constrained by the data, we still do not recover the shape of their radius valley.

Potential explanations include our difference in occurrence rate methodology, where F17 used the IDEM and did not take into account uncertainty in planet radius. In particular, they estimated that the underlying radius distribution after removing the smear due to this uncertainty would cause the gap to become slightly deeper, but the sub-Neptune peak would be increased (see Figure 7 of F17). Furthermore, while the CKS sample represented a significant improvement in precise stellar radii over previous works, it did not yet incorporate \textit{Gaia} DR2 parallaxes like the \citet{ber18} catalogue used here. \citet{ber18} compared histograms of planet radii (uncorrected for occurrence rates) using their catalogue and CKS-derived radii. Even under the same cuts as utilized by F17, they found a higher number of sub-Neptunes in their sample \cite[see Fig. 8 of ][]{ber18}.

\begin{figure}[t!]
\centering
\includegraphics[width=\linewidth]{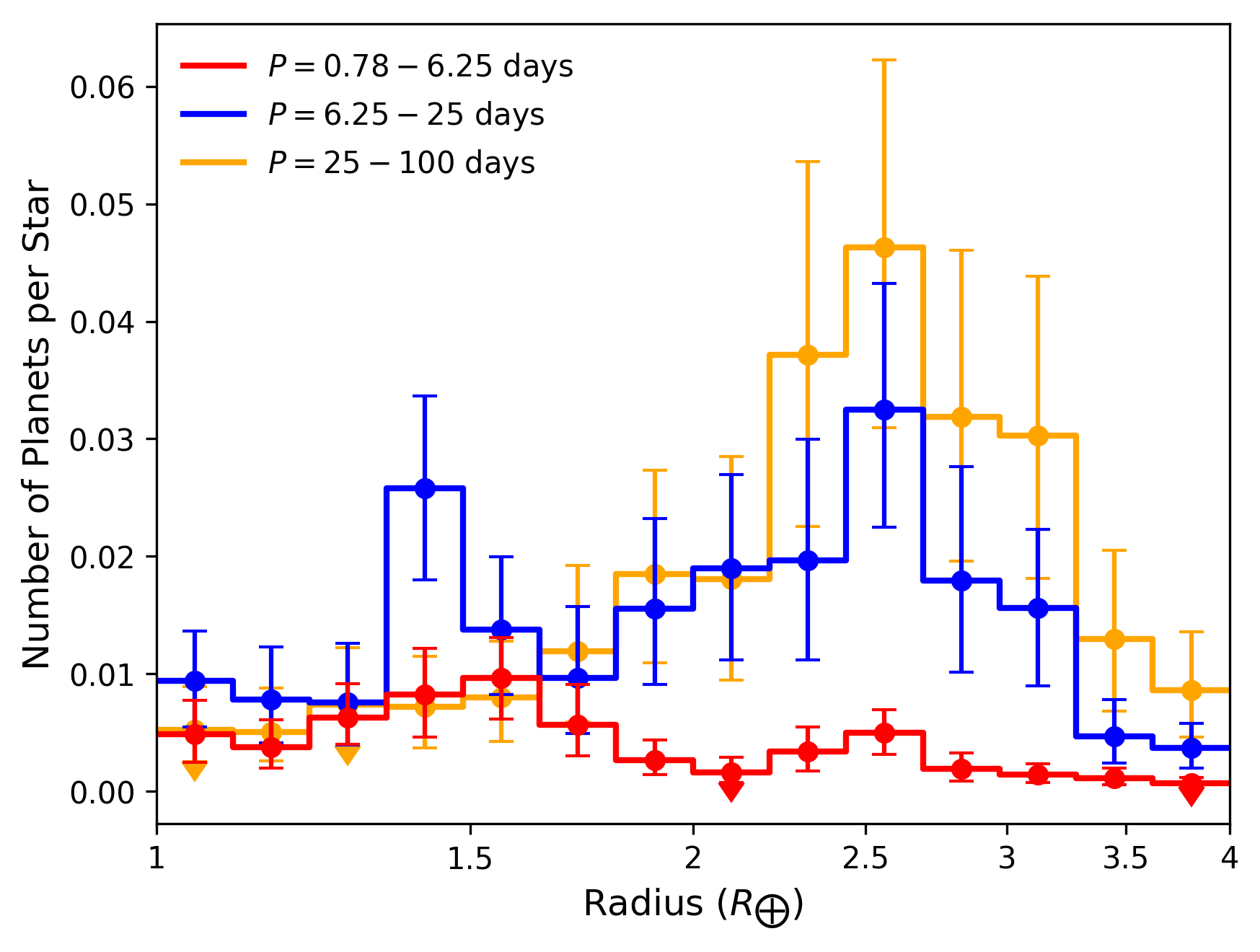}
\caption{Baseline FGK $P < 100$ day occurrence rates split over smaller period ranges to compare to the \citet{owe17} evolutionary model results.}\label{fig:valley_period}
\end{figure}

We continue our analysis with these caveats and maintain our focus on overall planet occurrence rates for FGK stars (using the full 96,280-star sample, without the magnitude cut). We have the means to investigate the radius valley further, as a function of period, due to finding separate occurrence rates over specific period bins. We show these results in Fig. \ref{fig:valley_period}. \citet{owe17} revisited the radius valley following the results of F17, and developed an analytical model to demonstrate that photoevaporation should separate planets into bare cores ($\sim1.3$ $R_{\bigoplus}$) and those with double the core's radius ($\sim2.6$ $R_{\bigoplus}$). Starting with a primordial \textit{Kepler} planet population and evolving the population under the effects of cooling contraction and mass loss by evaporation, \citet{owe17} produced a prediction for the final radius distribution across different period ranges. Within $P < 10$ days, they found that super-Earths dominate, with only a small peak past 2 $R_{\bigoplus}$. This is qualitatively similar to our $P < 6.25$ day occurrence rates. Not shown are our results for only $P < 3.13$ days, over which the sub-Neptune peak completely disappears. The $10-20$ day bins in \citet{owe17} demonstrated the most significant radius valley, with peaks at $\sim$1.3 $R_{\bigoplus}$ and $\sim$2.6 $R_{\bigoplus}$ exhibiting similar heights. Importantly, we clearly recover a similar bimodal distribution with strong peaks at $\sim$1.3 $R_{\bigoplus}$ and $\sim$2.6 $R_{\bigoplus}$ over roughly the same periods ($6.25 - 25$ days). Lastly, \citet{owe17} found that the distribution became dominated by sub-Neptunes at larger orbital periods, with only a small super-Earth peak over $20 - 40$ days and no such peak over $40 - 100$ days. We find that the super-Earth peak disappears earlier, with no such peak over $25 - 100$ days. Overall, our occurrence rates provide strong observational evidence in support of the \citet{owe17} model and can inform future studies on theoretical explanations for the radius valley.

\subsection{Dependence on Orbital Period}\label{sec:period}

\begin{figure}[t!]
\centering
\includegraphics[width=\linewidth]{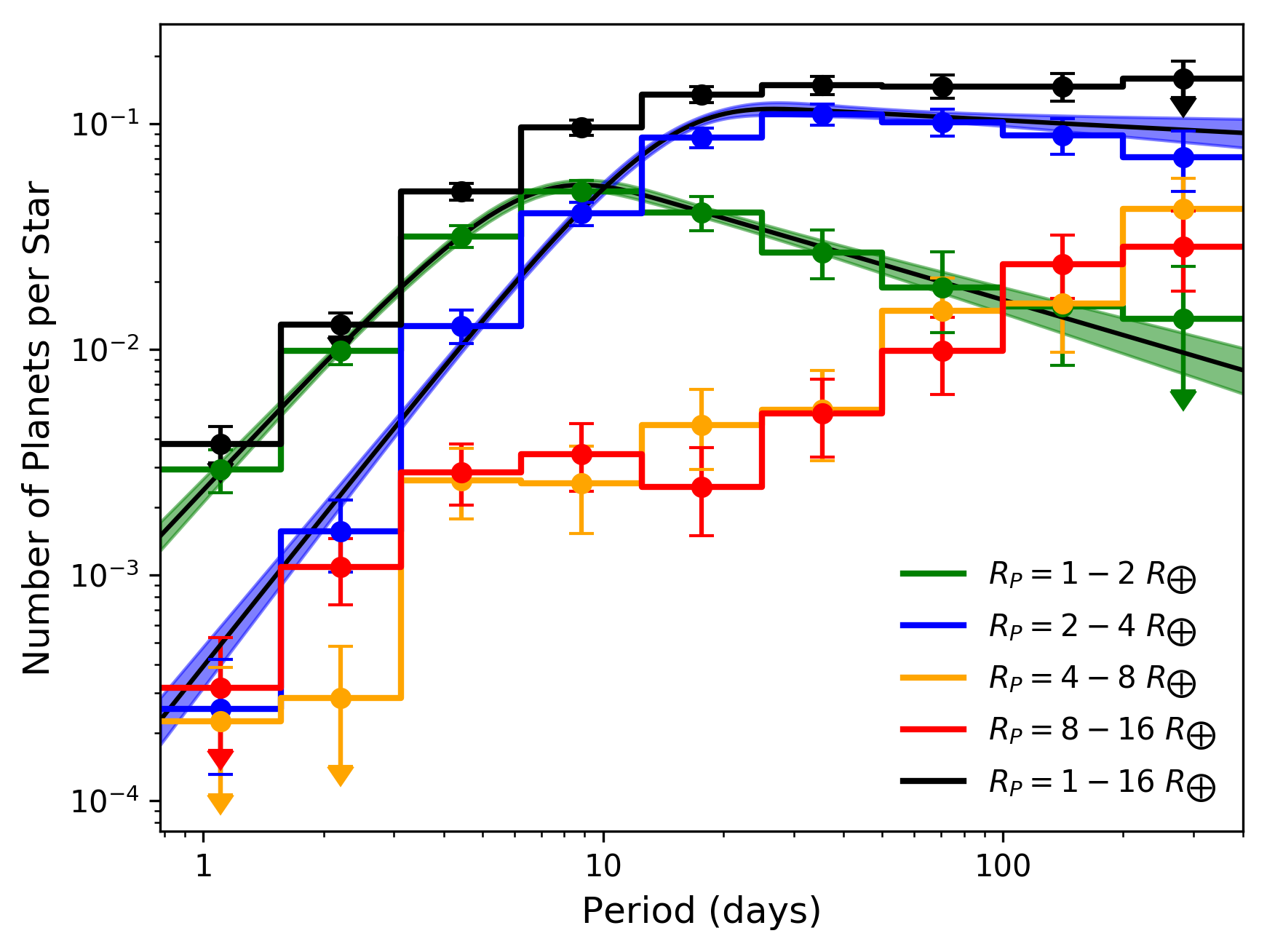}
\caption{FGK occurrence rates as a function of period, marginalized over different radius ranges. Also shown are fits of Eqn. \ref{eqn:period} to the $1-2$ $R_{\bigoplus}$ and $2-4$ $R_{\bigoplus}$ distributions.}\label{fig:period}
\end{figure}

Our FGK occurrence rate results marginalized over different radius ranges are shown in Fig. \ref{fig:period}. 

The $1-2$ $R_{\bigoplus}$ and $2-4$ $R_{\bigoplus}$ distributions show clear increases in $df/d\log{P}$ with $P$ up to a transition period $P_{0}$, followed by a decreasing trend for $1-2$ $R_{\bigoplus}$ and a flatter trend for $2-4$ $R_{\bigoplus}$. $P_{0}$ could indicate an important orbital distance down to which migration deposits planets, and differences between these distributions could indicate such mechanisms depend on planet size. We fit the function

\begin{equation}\label{eqn:period}
    F = \frac{df}{d\log{P}} = C P^{\beta} \big(1 - e^{-(P/P_{0})^{\gamma}}\big)
\end{equation}

\noindent from \citet{how12} to these distributions using the extension of their maximum-likelihood method outlined in \citet{pet18}. For the $i$th bin, the log-likelihood of the model is 

\begin{equation}
    \ln L_{i} = n_{\text{pl},i} \ln F \Delta \bm{x} + n_{\text{nd},i} \ln (1 - F \Delta \bm{x})
\end{equation}

\noindent where $\Delta \bm{x} = \Delta \log P \times \Delta \log R_{p}$ is the size of the bin, $n_{\text{pl},i}$ is the number of planets detected in the bin, and $n_{\text{nd},i} = n_{\text{pl},i}/f_{i} - n_{\text{pl},i}$ is the effective number of nondetections as estimated using the bin's occurrence rate $f_{i}$. The maximum-likelihood solution is obtained by maximizing the combined log-likelihood over all bins:

\begin{equation}
    \ln L = \sum_{i=1}^{n_{\text{bin}}} L_{i}.
\end{equation}

\noindent We used MCMC sampling with \texttt{emcee} to explore the parameter space. The median and 68.3\% credible interval for each distribution is shown in Eqn. \ref{eqn:period}, with associated parameters in Table \ref{tbl:period}.

\begin{table}[]
    \centering
    \caption{Median and 68.3\% credible interval parameters for Eqn. \ref{eqn:period}, describing the shape of the occurrence rate distribution with orbital period over different size ranges.}
    \label{tbl:period}
    \begin{tabular}{c|c|c|c|c}
    \hline
    \hline
        $R_{p}$ ($R_{\bigoplus}$) &  $C$ & $\beta$ & $\gamma$ & $P_{0}$ (days) \\
        \hline
        $1-2$ & $0.36_{-0.10}^{+0.07}$ & $-0.5_{-0.1}^{+0.1}$ & $2.42_{-0.1}^{+0.1}$ & $5.9_{-0.5}^{+0.5}$ \\
        $2-4$ & $0.33_{-0.12}^{+0.08}$ & $-0.1_{-0.1}^{+0.1}$ & $2.3_{-0.1}^{+0.1}$ & $13.3_{-1.5}^{+1.4}$
    \end{tabular}
\end{table}

This function simplifies to two power laws far from $P_{0}$:

\begin{equation}
    \frac{df}{d\log{P}} \propto
    \begin{cases}
        P^{\alpha} & \text{if }P \ll P_{0}\text{, where }\alpha = \beta + \gamma \\
        P^{\beta} & \text{if }P \gg P_{0}.
    \end{cases}
\end{equation}

\noindent The $1-2$ $R_{\bigoplus}$ planet occurrence rate rises with $P$, with $df \propto P^{\alpha} d\log{P}$ where $\alpha = 1.9_{-0.1}^{+0.1}$, up to a transition period $P_{0} = 5.9_{-0.5}^{+0.5}$ days. Beyond this, $df \propto P^{\beta} d\log{P}$ where $\beta = -0.5_{-0.1}^{+0.1}$. Comparatively, \citet{pet18} looked at $1-1.7$ $R_{\bigoplus}$ bins and found a slightly higher initial increase ($\alpha = 2.4_{-0.3}^{+0.4}$), a slightly higher transition period ($P_{0} = 6.5_{-1.2}^{+1.6}$ days), and a slightly shallower decrease at longer orbital periods ($\beta = -0.3_{-0.2}^{+0.2}$), though the 68.3\% credible intervals of the latter two parameters overlap with ours. Meanwhile, the long-period distribution found by \citet{don13} was flat, with $\beta = -0.10\pm0.12$.

The transition for $2-4$ $R_{\bigoplus}$ occurrence rates occurs farther out, at $13.3_{-1.5}^{+1.4}$ days, with a similar rapidly rising distribution at shorter orbital periods ($\alpha = 2.2_{-0.1}^{+0.1}$) and a nearly flat distribution at longer orbital periods ($\beta = -0.1_{-0.1}^{+0.1}$). These are consistent with a $P_{0} = 11.9_{-1.5}^{+1.7}$ day transition period, $\alpha = 2.3_{-0.2}^{+0.2}$, and $\beta = -0.1_{-0.1}^{+0.1}$ for $1.7-4$ $R_{\bigoplus}$ planets from \citet{pet18}. \citet{don13} also found a nearly flat distribution for $2-4$ $R_{\bigoplus}$ with $\beta = 0.11\pm0.05$. However, while all three studies agree that $\sim1-2$ $R_{\bigoplus}$ planets are more common than $\sim2-4$ $R_{\bigoplus}$ planets before the small-planet transition, our results and those of \citet{pet18} would indicate the opposite is true past the transition while \citet{don13} found similar occurrence rates for both distributions. We believe this is due to our use of more up-to-date and precise stellar radii, causing many $1-2$ $R_{\bigoplus}$ planets to be pushed into the $2-4$ $R_{\bigoplus}$ bin.

All three studies indicate that the distributions of larger planets (here, $4-8$ $R_{\bigoplus}$ and $8-16$ $R_{\bigoplus}$) are inconsistent with this power-law cut-off model, with occurrence rates gradually increasing over the entire period range. Our $4-8$ $R_{\bigoplus}$ occurrence rates do jump suddenly at 3.1 days, though this is likely another look at the Neptune Desert described in \S\ref{sec:radius}.

For $8-16$ $R_{\bigoplus}$, we do not confirm the three-day ``pile-up'' of hot Jupiters clear from radial velocity (RV) surveys \cite[e.g.][]{cum99, udr03, wri09}. This pile-up features strongly in various high-eccentricity migration scenarios \cite[e.g.][]{fab07, wu07, wu11}. However, other \textit{Kepler}-based studies have called into question the pile-up \citep{how12, fre13}, and differences in overall hot Jupiter ($P \lesssim 10$ days) occurrence rates between \textit{Kepler} and RV surveys have been previously noted. In particular, \textit{Kepler} hot Jupiter occurrence rates typically lie at around $0.4 - 0.6$\% \cite[e.g.][]{how12, fre13, mul15a, pet18} while RV occurrence rates are at around $0.9-1.2$\% \cite[e.g.][]{mar05, may11}. Our own hot Jupiter estimate should be intermediate between $0.43_{-0.09}^{+0.10}$\% ($0.78 - 6.25$ days) and $0.77_{-0.14}^{+0.16}$\% ($0.78 - 12.5$ days), consistent with other \textit{Kepler} results. \citet{daw13} suggested that the pile-up was a feature of metal-rich stars ([Fe/H]$ \geq$ 0) specifically, while the \textit{Kepler} sample has systematically lower metallicity than RV samples. They largely recovered the pile-up in the \textit{Kepler} sample when considering only stars with super-solar metallicity. Giant planet occurrence has also been shown to correlate strongly with host-star metallicity \citep{san03, fis05, pet18}. An assessment of the presence of the pile-up in our planet catalogue under these conditions will be left to a future paper focused on planet occurrence and its dependence on stellar metallicity.

\newpage

\subsection{Impact of Catalogue Reliability}\label{sec:reliability}

While our baseline results incorporate catalogue completeness, our planet sample is also not expected to be completely ``reliable.'' Signals not caused by planet transits, whether transit-like noise or astrophysical FPs, may be erroneously classified as planets by the vetting pipeline. The corresponding concept of reliability refers to the fraction of planets in the catalogue that are actually planets. We focus on our catalogue's reliability against noise specifically, which is the largest concern for candidates near our S/N = 6, three-transit detection limit, including small, rocky planets in orbits with long orbital periods.

The incorporation of reliability against noise into occurrence rate estimates remains an open question and was only first directly tackled in \citet{bry19}. \citet{bry19} took a probabilistic approach to reliability models, fitting components of the reliability with functions over finely spaced bins in period-S/N space (where S/N is represented by the MES employed by the \textit{Kepler} pipeline). Rather than assume a functional form of the reliability, we take a simplified approach using our reliability results from Paper I, summarized below.

We simulated noise using two datasets from the original 198,640 light curves searched. First, we recreated the Inverted (INV) set described in \citet{chr17} for their own reliability tests by inverting the light curves. Second, we recreated the Scrambled Group 1 (SCR1) set by reordering the \textit{Kepler} quarters according to the first order described in \citet{cou17}. Each of these datasets allowed for the existence of realistic signals with noise properties similar to the real data, while removing the possibility of planet transit detection. After searching and vetting this data, we estimated the fraction of noise FPs successfully classified as FPs (the ``effectiveness'' of the pipeline, $E$) to be 99.9$\%$ overall, having passed only 36 of 27,386 noise FPs as PCs. Using the definition of reliability from \citet{tho18},

\begin{equation}
    R = 1 - \frac{N_{\text{FP}}}{N_{\text{PC}}}\bigg(\frac{1-E}{E}\bigg),
\end{equation}

\noindent where $N_{\text{PC}}$ and $N_{\text{FP}}$ are the number of observed PCs and FPs identified by the vetting pipeline, our pipeline has an overall reliability against noise FPs of 98.3$\%$.

\begin{figure*}[ht!]
\centering
\includegraphics[width=0.8\textwidth]{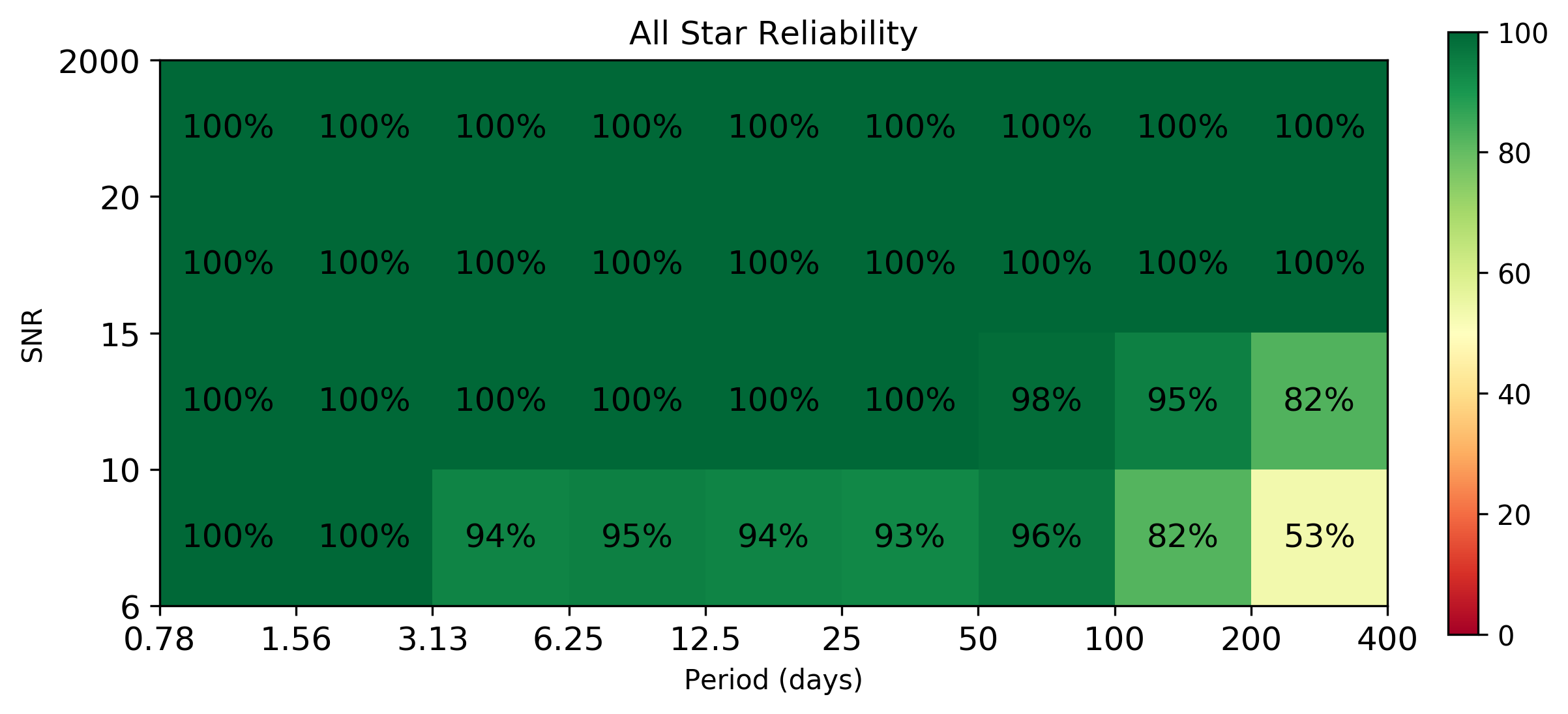}
\includegraphics[width=0.8\textwidth]{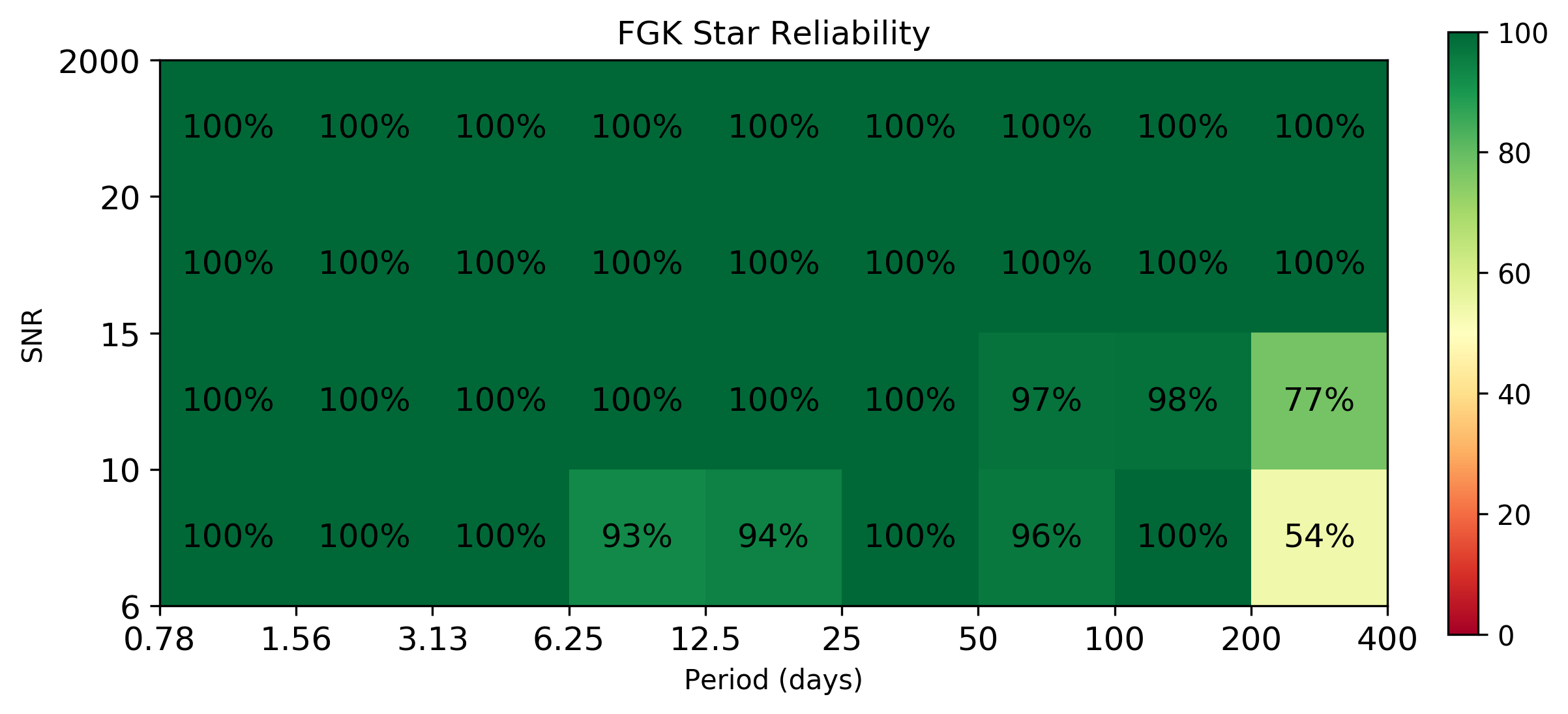}
\caption{Estimate of the reliability of our catalogue, where reliability refers to the fraction of PCs within a given period-S/N bin that are actually planets. Top: considering all stars searched in the \textit{Kepler} sample. Bottom: only including the FGK stars used in this study.}\label{fig:reliability}
\end{figure*}

To set up the application of these results to our occurrence rate estimates, we determined reliability over a coarse grid in period-S/N space in an attempt to reduce the effect of small number statistics. We chose period bin edges equal to those used for our occurrence rates and make the assumption that the reliability of each bin is constant across that bin. While ideally we would find our reliability across only the FGK stars in our sample, a concern is that small number statistics would be more significant than for the full, $\sim$200,000-star sample. Thus, while we recognize that noise properties between the two samples should be different, we elected to use our results across all stars (top panel of Fig. \ref{fig:reliability}) to improve the signal-to-noise ratio of the estimate. Our FGK-only results are included in the bottom panel of Fig. \ref{fig:reliability} for comparison.

Lastly, to incorporate reliability into the ABC methodology itself, we recall the discussion of selection effects from \citet{you11} as outlined in \S\ref{sec:forwardmodel}. The FP-related selection effect was $\eta_{\text{fp}} = 1/(1 - r_{\text{fp}})$, where $r_{\text{fp}}$ is the rate of FP events that are detected as planets. This is equal to 1 minus the reliability, giving $\eta_{\text{fp}} = 1/R$. Thus, we replace Eqn. \ref{eqn:net} in our forward model with

\begin{equation}
    \eta_{\text{tot}} = \eta_{\text{tr}}\eta_{\text{rec}}/R,
\end{equation}

\noindent where $R$ is found for a given planet according to its period and S/N.

We recalculated our FGK occurrence rates under these changes, with median and 68.3\% credible interval results included in Table \ref{tbl:FGKrates} alongside our baseline estimates. Reliability did not significantly impact our results, with posteriors for every cell overlapping significantly. This is not unexpected, given that our reliability is high ($> 90$\%) outside only the long-period, low-S/N corner in period-S/N space. For the corresponding period-radius cells, it is difficult to assess the full impact of reliability given the same areas have very low completeness and no planet detections, meaning estimates are already not as well constrained as other areas. However, we do tend to reduce our upper limits in the most affected areas. In particular, our $0.5 - 0.71$ $R_{\bigoplus}$, $200 - 400$ day cell had its upper limit reduced from 14.4\% to 13.1\%, and our $0.71 - 1$ $R_{\bigoplus}$, $200 - 400$ day cell had its upper limit reduced from 3.4\% to 2.7\%.

Given that we focused in previous sections on high-reliability regimes (i.e. mainly $R_{p} > 1$ $R_{\bigoplus}$, $P < 200$ days), we do not expect the lack of reliability incorporation to affect our previous analysis. However, for our upcoming terrestrial HZ planet frequency discussion, which will specifically depend on calculations over regions of low reliability, we will report both versions of occurrence rate results.

\section{Terrestrial HZ Planet Frequency}\label{sec:etaearth}

A partial explanation for the lack of consistency between literature $\eta_{\bigoplus}$ values lies in how authors define the ``habitable zone.'' In a landmark study, \citet{kas93} placed estimates of the boundaries of the HZ using one-dimensional, cloud-free climate models. According to these models, the runaway greenhouse effect places an inner edge at 0.84 AU, while a more conservative estimate places the inner edge due to water loss at $0.95$ AU. Meanwhile, an outer edge at $1.67$ AU is determined from the maximum greenhouse effect from a CO$_{2}$ atmosphere. \citet{kop13} revised the \citet{kas93} estimates with an updated climate model, which moved the water-loss inner edge up to 0.99 AU and the maximum greenhouse limit out to 1.70 AU. \citet{kop13} also produced a wider HZ range based on the flux received by recent Venus (0.75 AU) and early Mars (1.77) AU. These two ranges ($0.99 - 1.70$ AU and $0.75 - 1.77$ AU) are referred to as ``conservative'' and ``optimistic'' HZs respectively. Many occurrence rate papers including \citet{sil15}, \citet{gar18}, \citet{zin19b}, \citet{bry19}, and \citet{hsu19} have incorporated these definitions. Other employed definitions include an incident flux range within a factor of 4 from that received by the Earth \citep{pet13}.

Another complicating factor is how authors define the size of a potentially habitable, rocky planet. Too small, and a planet will not be able to retain an atmosphere or support plate tectonics \citep{kas93}. \citet{ray07}, for instance, considered 0.3 $M_{\bigoplus}$ as the lower-mass limit for planetary habitability. Using the mass-radius relation for $R_{p} < 1.23$ $R_{\bigoplus}$ from \citet{che17},

\begin{equation}
    M_{p} = 0.972\bigg(\frac{R_{p}}{R_{\bigoplus}}\bigg)^{3.584} M_{\bigoplus},
\end{equation}

\noindent this corresponds to a radius of 0.72 $R_{\bigoplus}$. This lower limit was used by \citet{zin19b}, while other studies have somewhat arbitrarily used lower bounds anywhere between 0.5 and 1 $R_{\bigoplus}$. As for an upper radius limit, we must consider a potential transition between rocky super-Earths and volatile-shrouded sub-Neptunes. It is difficult to simplify this to a single radius as the composition of a planet is much more informative about its potentially rocky nature. However, \citet{rog15} took a statistical approach to a sample of small planets with both masses and radii, finding that most planets above 1.6 $R_{\bigoplus}$ are not expected to be rocky, and a best-fit transition occurs at $R_{p} = 1.48_{-0.04}^{+0.08}$. Furthermore, as found in \citet{ful17} and further explored here, $\sim$1.5 $R_{\bigoplus}$ precedes a gap in the exoplanet size distribution, which would support this prediction. Previous papers have typically chosen upper radius limits between $1.5$ and $2$ $R_{\bigoplus}$.

In an effort to standardize $\eta_{\bigoplus}$ determination, the ExoPAG SAG13 report recommended that authors produce estimates using both $1 - 1.5$ $R_{\bigoplus}$ and $0.5 - 1.5$ $R_{\bigoplus}$ radius ranges. They also defined their G-type star ``$\eta_{\text{habSol,SAG13}}$'' value as lying between 237 and 860 days, corresponding to the \citet{kop13} optimistic HZ.

Following \citet{hsu19}, we start with occurrence rates using bin edges of $R_{p} = \{0.5, 0.75, 1, 1.25, 1.5, 1.75, \newline 2\}$ $R_{\bigoplus}$ and $P = 237 - 500$ days. These are the same radius bin edges as in \citet{hsu19}, but with an additional $0.5 - 0.75$ $R_{\bigoplus}$ bin to meet the radius range recommendation of SAG13. Meanwhile, the 237 day lower bound on the period corresponds to the inner edge of the optimistic HZ as defined by \citet{kop13} for a Sun-like star. The 500 day upper bound corresponds to the limit of \textit{Kepler}'s (and our pipeline's) sensitivity. As in the SAG13 report, we considered G-type stars to represent ``Sun-like'' stars for our calculations. SAG13 defined G-type stars using the same temperature limits as we use here ($5300 - 6000$ K).

Because we are also interested in incorporating reliability, we needed to find our catalogue's reliability specific to the $237 - 500$ day period range. Using our all-star results from \S\ref{sec:reliability}, we found $R = 0.39$, $0.80$, and $1.0$ for S/N $<$ 10, 10 $\leq$ S/N $<$ 15, and S/N $\geq$ 15.

\subsection{Optimistic HZ Estimate}

Our direct calculation over the $237 - 500$ day bin represents a subset of the $237 - 860$ day optimistic HZ. For the $1 - 1.5$ $R_{\bigoplus}$ range, we find an occurrence rate of $0.05_{-0.03}^{+0.04}$ planets per star. When incorporating reliability, we find an occurrence rate of $0.05_{-0.02}^{+0.03}$ planets per star. For the $0.75 - 1.5$ $R_{\bigoplus}$ range considered by \citet{hsu19}, we find an occurrence rate of $0.12_{-0.06}^{+0.08}$ ($0.10_{-0.04}^{+0.07}$ with reliability), which well overlaps with the $0.16_{-0.06}^{+0.11}$ estimate from \citet{hsu19}. Lastly, we find a considerably increased estimate with substantial uncertainties for $0.5 - 1.5$ $R_{\bigoplus}$, at $0.38_{-0.19}^{+0.29}$ ($0.31_{-0.15}^{+0.28}$), on account of the low completeness and poor constraints provided by the data. Note that neither of our studies had planet detections over any of these radius ranges, so these occurrence rates are best interpreted as upper limits. 

Considering the entire $237 - 860$ day HZ range requires extrapolating these results to longer periods. Interpreting results obtained via extrapolation should be done with added caution, considering we have demonstrated substantial uncertainties in sub-$\eta_{\bigoplus}$ occurrence rate estimates and will necessarily have to make an assumption about the nature of planet distributions beyond the limit of our sensitivity. However, the ABC methodology requiring such extrapolation is not unique. All studies are limited by the \textit{Kepler} mission duration to planets within $\approx 500$ days, and small planets typically require more observed transits than larger planets in order to produce signals with sufficient S/N for detection. Other grid-based occurrence rates must make similar assumptions to our work. Meanwhile, studies that find a function to describe planet distributions with period and/or radius may integrate over a desired $\eta_{\bigoplus}$ range to produce an estimate, but the function itself will have been based on planets with larger sizes and/or shorter orbital periods. In these cases, the employed assumption is that the same model can also explain the $\eta_{\bigoplus}$ regime, which is not necessarily true; \citet{mul18}, for instance, found that their broken power-law model broke down outside of 400 days. Furthermore, we have shown that there are numerous period- and size-dependent small-scale variations (such as the clear radius valley for orbital periods within 25 days), indicating that a parametric occurrence rate model is not necessarily the best descriptor of the data even over the shorter orbital periods and larger planet sizes considered by these works.

With these caveats, we adopt the method of extrapolation used by \citet{hsu19} and assume that the differential occurrence rate derived over $237 - 500$ days and a given radius range is the same over longer periods. Under this assumption, we estimate optimistic HZ occurrence rates of $\eta_{\bigoplus} = 0.08_{-0.04}^{+0.07}$ ($0.08_{-0.04}^{+0.06}$) planets per star for $1 - 1.5$ $R_{\bigoplus}$, $\eta_{\bigoplus} = 0.21_{-0.10}^{+0.14}$ ($0.17_{-0.08}^{+0.11}$) for $0.75 - 1.5$ $R_{\bigoplus}$, and $\eta_{\bigoplus} = 0.66_{-0.32}^{+0.51}$ ($0.53_{-0.26}^{+0.48}$) for $0.5 - 1.5$ $R_{\bigoplus}$.

\subsection{Conservative HZ Estimate}

The $0.99 - 1.70$ AU conservative HZ from \citet{kop13} corresponds to orbital periods of $360 - 809$ days. Extrapolating our $237 - 500$ day results over these periods, we find occurrence rates of $\eta_{\bigoplus}$ = $0.05_{-0.03}^{+0.04}$ ($0.05_{-0.02}^{+0.04}$) planets per star for $1 - 1.5$ $R_{\bigoplus}$, $\eta_{\bigoplus}$ = $0.13_{-0.06}^{+0.09}$ ($0.11_{-0.05}^{+0.07}$) for $0.75 - 1.5$ $R_{\bigoplus}$, and $\eta_{\bigoplus}$ = $0.42_{-0.20}^{+0.32}$ ($0.34_{-0.16}^{+0.30}$) for $0.5 - 1.5$ $R_{\bigoplus}$.

\subsection{Comparison to Previous Works}

The challenges involved with defining the bounds of $\eta_{\bigoplus}$ have motivated recent studies to instead report and compare $\Gamma_{\bigoplus}$, the differential occurrence rate near the HZ \citep{you11, for14, bur15}. We follow \citet{hsu19} in defining $\Gamma_{\bigoplus}$ using our $237 - 500$ day, $0.75 - 1.5$ $R_{\bigoplus}$ results, giving $\Gamma_{\bigoplus} \equiv (d^{2}f)/[d(\ln{P})d(\ln{R_{p}})] = 0.23_{-0.11}^{+0.16} (0.19_{-0.08}^{+0.13})$.

\begin{figure}[t!]
\centering
\includegraphics[width=\linewidth]{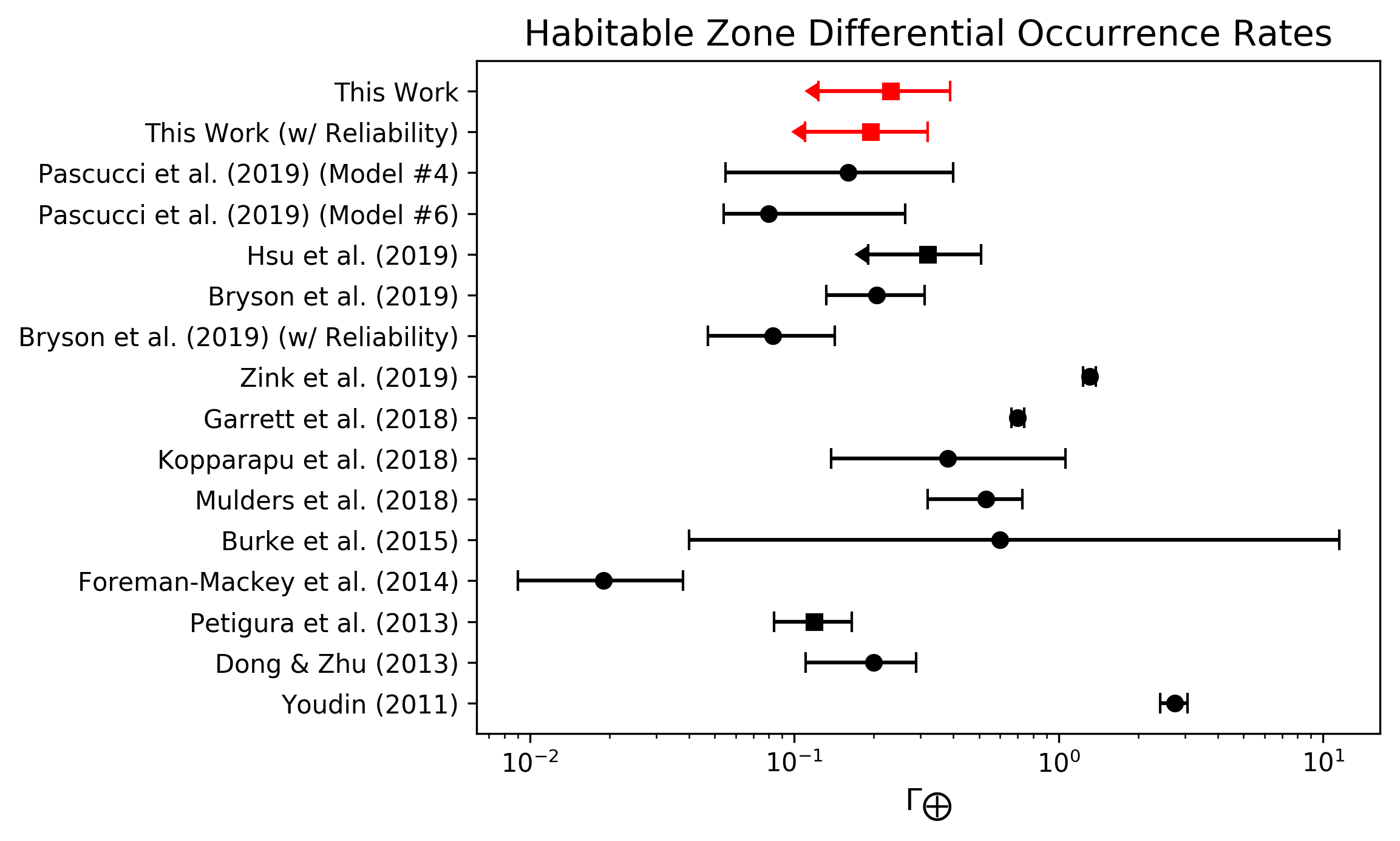}
\caption{A collection of $\Gamma_{\bigoplus}$ values from the literature: \citet{pas19}, \citet{hsu19}, \citet{bry19}, \citet{zin19a}, \citet{gar18}, ExoPAG SAG13 via \citet{kop18}, \citet{mul18}, \citet{bur15}, \citet{for14}, \citet{pet13}, \citet{don13}, and \citet{you11}. Squares indicate that grid-based occurrence rates were explored (our work, \citet{hsu19}, and \citet{pet13}), while circles indicate a functional form for the occurrence rate was assumed (all others). Left-pointing arrows indicate that the result is meant to be interpreted as an upper limit.}\label{fig:gamma}
\end{figure}

Comparisons with other $\Gamma_{\bigoplus}$ estimates in the literature are shown in Fig. \ref{fig:gamma}. The values from \citet{pas19} correspond to their Model \#4 and \#6 results given in their Table 2, which address the impact of photoevaporated cores by excluding planets within 12 and 25 days from their analysis, respectively. The ExoPAG SAG13 estimate, obtained via \citet{kop18}, is based on a meta-analysis of community-submitted G-type star occurrence rate studies, for which a broken power law was fit to a combined period-radius grid. The plotted central values were found by plugging in the Earth's radius and period into their baseline power law, while the lower and upper uncertainties correspond to their pessimistic and optimistic power laws, respectively. The values from \citet{bur15} represent the allowable range from their sensitivity analysis. The \citet{pet13} result is calculated by converting their $200 - 400$ day, $1 - 2$ $R_{\bigoplus}$ extrapolated occurrence rate into a differential rate. The \citet{don13} value is an extrapolation of their $1 - 2$ $R_{\bigoplus}$ best-fit function, evaluated at the orbital period of Earth and with error bars given by the propagation of uncertainty. All other values are the $\Gamma_{\bigoplus}$ results explicitly reported in their respective papers.

Our results compare most favourably to those from \citet{pas19}, \citet{hsu19}, \citet{bry19}, \citet{kop18}, and \citet{don13}, and are well within the allowable range of \citet{bur15}. Meanwhile, the values from \citet{you11}, \citet{gar18}, and \citet{zin19a} all lie above our upper limits. Lack of consistency with the early \citet{you11} result can readily be explained by the fact that it is based on a much older \textit{Kepler} catalogue, having included only planets with $P < 50$ days. Part of the disagreement with \citet{zin19a} can be explained by their incorporation of transit multiplicity --- in other words, they took into account a reduction in detection efficiency with planet detection order, which would result in an increased occurrence rate --- though this is unlikely to explain the whole discrepancy (see \S\ref{sec:limitations}). Nevertheless, we argue that our uncertainty estimates are more realistic than those of \citet{gar18} and \citet{zin19a}.

Turning to the impact of reliability, we again do not find significantly different occurrence rates. This is contrasted to the results of \citet{bry19} who found that incorporating reliability dropped $\Gamma_{\bigoplus}$ from $0.205_{-0.073}^{+0.106}$ down to $0.083_{-0.036}^{+0.050}$). We note three things: first, we reemphasize that given the total lack of planet detections in the relevant bins, our results are poorly constrained by the data and should be interpreted as upper limits. Had we been able to place better constraints on these occurrence rates, a clearer picture of the full impact of reliability would appear. Second, we only took into account reliability against noise and systematics, whereas \citet{bry19} additionally considered reliability against astrophysical FPs. Their false-alarm-only reliability was $\Gamma_{\bigoplus} = 0.128_{-0.049}^{+0.077}$. Lastly, our grid-based occurrence rate method means that the reliability of a given bin will only directly affect the occurrence rates of that bin, and only indirectly affect the occurrence rate of neighbouring, simultaneously fit bins due to the leaking of planets between bins. \citet{bry19} fit a joint power-law model across all planets with $R_{p} = 0.75 - 2.5$ $R_{\bigoplus}$ and $P = 50 - 400$ days, meaning that the reliability of any planet would affect the fit results across the entire space considered. We would expect this to cause the reliability to have a greater affect on occurrence rates than in our work.

A subset of $\eta_{\bigoplus}$ not yet mentioned is the concept of $\zeta_{\bigoplus}$: the number of planets per star with radii and orbital periods within 20$\%$ of Earth's values, as introduced by \citet{bur15}. Combining both an extrapolated analysis (assuming their $0.75 - 2.5$ $R_{\bigoplus}$, $50 - 300$ day broken power law) and a direct analysis (recomputing a broken power law over $300 - 700$ days), \citet{bur15} reported $\zeta_{\bigoplus} = 0.10$ planets per star with an allowable range of $0.01- 2$. To compare to this value, we recalculated our occurrence rates with bin edges of $R_{p} = \{0.8, 1.0, 1.2\}$ $R_{\bigoplus}$ and $P = 292 - 438$ days. We recovered their central value almost exactly, finding $\zeta_{\bigoplus} = 0.12_{-0.06}^{+0.09}$ ($0.10_{-0.06}^{+0.07}$).

\section{Limitations of Our Occurrence Rates}\label{sec:limitations}

Here, we summarize the main assumptions and design choices made in our study, each of which can motivate future improvements to our occurrence rate model and methodology.

\textit{Astrophysical FPs}: We assumed that the FP rate due to astrophysical FPs (such as background, grazing, or hierarchical triple eclipsing binaries) was negligible. In order to minimize the effect of contamination from these FPs on our occurrence rates, we vetted against them in our Paper I pipeline and only included candidates in our final catalogue that were also passed as PCs by the \textit{Kepler} Robovetter. However, in the future we could characterize our catalogue's reliability against astrophysical FPs similarly to how we explored reliability against noise and systematics.

\textit{Behaviour of simulated noise}: We used the INV and SCR1 datasets to estimate our catalogue reliability, under the assumption that they accurately simulate the noise and systematics present in the actual observed \textit{Kepler} data. However, this is not the case for all periods and all types of false alarms. For example, the use of the INV set relies on the assumption that the false alarms are symmetric upon data inversion. This means it will not reproduce the drops in flux caused by cosmic-ray-induced sudden pixel sensitivity dropout \citep{jen10}. Meanwhile, the SCR1 dataset leaves each \textit{Kepler} quarter untouched. Significant signals with periods less than a month may appear in both the original and ``simulated'' noise sets.

\textit{Reliability}: We used a simplistic estimate of our catalogue's reliability against noise, in the form of a coarse grid in period-S/N space. \citet{bry19} found a functional for reliability, which we could adopt after a more thorough investigation into our catalogue's reliability against both astrophysical FPs and false alarms.

\textit{Choice of prior}: We assumed independent, uniform priors for all bins throughout this study. While this was the same choice used by \citet{hsu19} for their baseline study, they cautioned that this was equivalent to assuming a prior on the total rate that was peaked toward high occurrence rates. After using a Dirichlet prior over radius bins and a uniform prior on the total rate, they found significantly lower rates for bins poorly constrained by the data, such as the $\eta_{\bigoplus}$ regime. Currently, our ABC code is not able to implement a Dirichlet or other multivariate prior, though this would be a line of inquiry we would be interested in exploring in the future.

\textit{Eccentricity}: In our forward model, we simulated all planets in perfectly circular orbits ($e = 0$) for simplicity. Nonzero eccentricity would affect both the transit duration and detection probability of a given planet. \citet{bur15} found that introducing eccentricity had only a minor effect on lowering occurrence rates, comparable to systematic errors. However, in principle we could draw from a distribution for the eccentricity \cite[e.g.][]{hsu18, hsu19}, or take into account some dependence on planet properties in the future.

\textit{Planetary system architectures}: We have only attempted to characterize the average occurrence rates of individual planets, rather than the orbital architectures of multiplanet systems. Notably, \citet{mul18} introduced a forward model that simulates a planetary system for each star and takes into account correlations in the properties of each planet in the system (i.e. orbital inclination, period, and radius). However, their simulations were made possible by assuming a parametric function for exoplanet occurrence rates. Grid-based occurrence rates as in our study are more computationally limited given that we can only simulate $\sim$5 bins (covering a small section of period-radius space) at a time. Nevertheless, future improvements to the efficiency of ABC algorithms may make this possible.

\textit{Window function}: We adopted the binomial approximation of the window function probability of detecting at least three transits in \textit{Kepler} data from \citet{bur15}. This form of the window probability was also used in \citet{hsu18} and \citet{zin19a}. An alternative would be to use the \textit{Kepler} DR25 target-by-target window functions from \citet{bur17}, which have been shown to result in reduced occurrence rates due to better accounting for the detection probability for planets with few transits \citep{hsu19}. We chose not to use these data products given they are unique to the \textit{Kepler} DR25 pipeline and TPS algorithm, though given we use the same three-transit minimum detection criteria, the differences between our pipelines may be minor enough that it would be worth incorporating these data products in the future.

\textit{Transit multiplicity}: We did not take into account reductions in detection efficiency due to transit multiplicity. Similar to the \textit{Kepler} pipeline, our search was a multipassthrough process in which the strongest S/N signal in the light curve would be removed after detection in order to facilitate another search for more planets. With more potential signals removed from the data, subsequent searches would be based on less available data, and the detection probability of finding a surviving candidate would be reduced. \citet{zin19a} found that the \textit{Kepler} pipeline experiences an additional 5.5\% and 15.9\% efficiency loss for planets with $P < 200$ and $P > 200$ days, respectively, after finding the strongest transit signal in a multiple-planet system.

\subsection{Final $\eta_{\bigoplus}$ Recommendation}

For the definition of the HZ, \citet{kop13} recommended the use of their conservative ($0.99 - 1.70$ AU) boundaries. For the lower radius limit, we are inclined to consider potentially habitable rocky planets as those down to 0.75 $R_{\bigoplus}$ rather than 0.5 $R_{\bigoplus}$, as this is near the scientifically motivated 0.72 $R_{\bigoplus}$ limit used by \citet{zin19b}. For the upper radius limit, the 1.5 $R_{\bigoplus}$ radius considered throughout this section is already well motivated by both the characteristics of the radius valley and the findings of \citet{rog15}. Meanwhile, incorporating reliability allows for better constraints on $\eta_{\bigoplus}$ upper limits. 

For our reliability-incorporated, $0.75 - 1.5$ $R_{\bigoplus}$ conservative HZ, our $\{5, 15.9, 50, 84.1, 95\}$th percentiles are $\eta_{\bigoplus} = \{0.04,0.06,0.11,0.18,0.24\}$ planets per star. Thus, we suggest future exoplanet characterization and habitability-related missions to consider an upper limit (84.1th percentile) of $<0.18$ terrestrial HZ planets per Sun-like star.

Many of the limitations discussed in this section (reliability against astrophysical FPs, nonzero eccentricity, use of a Dirichlet prior, use of DR25 window functions) would lead to reduced occurrence rates should we apply them, so this upper limit should be robust to these concerns. However, transit multiplicity would indicate that our reported occurrence rates are underestimated. \citet{hsu19} estimated a less than $\sim$8\% increase in their DR25 occurrence rates for long-period planets due to this effect. Our pipeline may be slightly more affected given its lower recovery efficiency for planets with few transits. Thus, we estimate that our $\eta_{\bigoplus}$ upper limit is robust to $\sim$10\%.

\section{Conclusions}\label{sec:conclusions}

We have presented exoplanet occurrence rates for FGK-, F-, G-, and K- type stars, using approximate Bayesian computation --- a promising methodology that was only recently applied to exoplanet occurrence rates for the first time \citep{hsu18}. We further based our estimates on an independent analysis of the \textit{Kepler} light curves, produced by searching the entire \textit{Kepler} sample for known and new exoplanets with our own search and vetting pipeline. We directly incorporated search completeness, vetting completeness, and planet radius uncertainty. We also described a first step toward incorporating catalogue reliability into our occurrence rate measurements, and suggest that the impact of low reliability is less severe than for methods that assume parametric planet distribution functions.

In our investigation of the dependence of planet occurrence on host star effective temperature, we confirmed the findings of \citet{how12} and \citet{mul15a, mul15b} that small planets are significantly more abundant around cooler stars than hotter stars. We also provided new observational evidence for the radius gap and took a deeper look into its dependence on orbital period, finding strong agreement with recent model predictions \citep{owe17}. In our investigation of the overall distribution of planets with orbital period, we found that power laws well describe $1 - 2$ $R_{\bigoplus}$ and $2 - 4$ $R_{\bigoplus}$ occurrence rates, broken by transition periods of $5.9_{-0.5}^{+0.5}$ days and $13.3_{-1.5}^{+1.4}$ days respectively. Meanwhile, larger planets demonstrate a consistent rise in occurrence rates with orbital period across the whole $P < 400$ days examined.

Lastly, determining the frequency of potentially habitable planets was a primary motivation of our work. Given the sensitivity of $\eta_{\bigoplus}$ to the assumed definitions of the habitable zone and the size limits of potentially habitable, rocky planets, we reported a wide variety of upper limits that can be compared to other literature values, both before and after taking into account catalogue reliability. In conclusion, we recommend an upper limit of $<$0.18 potentially habitable planets per Sun-like star. Upon consideration of the assumptions made for our occurrence rate calculations, this upper limit should be robust to $\sim$10\%.

\section{Acknowledgements}

We thank NASA for providing the wealth of \textit{Kepler} data available to the public for download and analysis, without which this paper would not be possible. We also thank the referee for their invaluable comments and insight.

This work has made use of data from the European Space Agency (ESA) mission {\it Gaia} (\url{https://www.cosmos.esa.int/gaia}), processed by the {\it Gaia} Data Processing and Analysis Consortium (DPAC, \url{https://www.cosmos.esa.int/web/gaia/dpac/consortium}). Funding for the DPAC has been provided by national institutions, in particular the institutions participating in the {\it Gaia} Multilateral Agreement.

\facilities{Gaia, Kepler}

\software{\texttt{emcee}~\citep{for13}, \texttt{matplotlib}~\citep{hun07}, \texttt{numpy} \citep{vdw11}, \texttt{scipy} \citep{jon01}, \texttt{cosmoabc}~\citep{ish15}} 

\appendix

\section{Tables of Occurrence Rate Results}

Table \ref{tbl:FGKrates} gives our occurrence rate results for FGK-, F-, G-, and K-type stars over the baseline period-radius grid. The median and 68.3\% credible interval of each ABC posterior is shown. For bins with zero planet detections, we report only the upper limit (84.1th percentile). Note that ``w/ R'' indicates that we incorporated reliability against noise and systematics.

\begin{longtable*}[h!]{c|c|c|c|c|c|c}
\caption{Occurrence rate results for FGK-, F-, G-, and K-type stars over the whole period-radius grid. The ``FGK w/ $R$'' column refers to our FGK results after incorporating reliability against transit-like noise. Results are given in \% ($10^{-2}$).}\label{tbl:FGKrates}\\
\hline\hline
Period (days) & Radius ($R_{\bigoplus}$) & FGK (\%) & F (\%) & G (\%) & K (\%) & FGK w/ $R$ (\%) \\
\hline
\endfirsthead
Period (days) & Radius ($R_{\bigoplus}$) & FGK (\%) & F (\%) & G (\%) & K (\%) & FGK w/ $R$ (\%) \\
\hline
\endhead
$0.78 - 1.56$ & $0.50 - 0.71$ & $0.02_{-0.01}^{+0.02}$ & $<0.03$ & $0.03_{-0.02}^{+0.04}$ & $0.1_{-0.06}^{+0.09}$ & $0.02_{-0.01}^{+0.02}$ \\
$0.78 - 1.56$ & $0.71 - 1.00$ & $0.1_{-0.03}^{+0.04}$ & $0.02_{-0.01}^{+0.02}$ & $0.14_{-0.06}^{+0.07}$ & $0.3_{-0.14}^{+0.17}$ & $0.09_{-0.04}^{+0.04}$ \\
$0.78 - 1.56$ & $1.00 - 1.41$ & $0.18_{-0.05}^{+0.05}$ & $0.04_{-0.02}^{+0.03}$ & $0.23_{-0.08}^{+0.1}$ & $0.61_{-0.21}^{+0.23}$ & $0.18_{-0.05}^{+0.05}$ \\
$0.78 - 1.56$ & $1.41 - 2.00$ & $0.12_{-0.04}^{+0.04}$ & $0.03_{-0.02}^{+0.03}$ & $0.13_{-0.06}^{+0.07}$ & $0.4_{-0.18}^{+0.19}$ & $0.11_{-0.03}^{+0.04}$ \\
$0.78 - 1.56$ & $2.00 - 2.83$ & $0.01_{-0.01}^{+0.01}$ & $0.02_{-0.01}^{+0.02}$ & $<0.05$ & $<0.17$ & $0.01_{-0.01}^{+0.01}$ \\
$0.78 - 1.56$ & $2.83 - 4.00$ & $0.01_{-0.01}^{+0.01}$ & $0.01_{-0.01}^{+0.01}$ & $0.02_{-0.02}^{+0.02}$ & $<0.11$ & $0.01_{-0.01}^{+0.01}$ \\
$0.78 - 1.56$ & $4.00 - 5.66$ & $0.01_{-0.01}^{+0.01}$ & $<0.03$ & $0.02_{-0.02}^{+0.03}$ & $<0.11$ & $0.01_{-0.01}^{+0.01}$ \\
$0.78 - 1.56$ & $5.66 - 8.00$ & $<0.02$ & $<0.03$ & $<0.04$ & $<0.1$ & $<0.02$ \\
$0.78 - 1.56$ & $8.00 - 11.31$ & $<0.02$ & $<0.03$ & $<0.03$ & $<0.12$ & $<0.03$ \\
$0.78 - 1.56$ & $11.31 - 16.00$ & $0.02_{-0.01}^{+0.02}$ & $0.02_{-0.01}^{+0.02}$ & $0.02_{-0.02}^{+0.03}$ & $0.08_{-0.05}^{+0.08}$ & $0.02_{-0.01}^{+0.02}$ \\
\hline
$1.56 - 3.13$ & $0.50 - 0.71$ & $0.1_{-0.05}^{+0.06}$ & $0.03_{-0.02}^{+0.04}$ & $0.17_{-0.1}^{+0.12}$ & $0.22_{-0.14}^{+0.22}$ & $0.1_{-0.06}^{+0.06}$ \\
$1.56 - 3.13$ & $0.71 - 1.00$ & $0.25_{-0.08}^{+0.09}$ & $0.08_{-0.05}^{+0.07}$ & $0.33_{-0.14}^{+0.16}$ & $0.66_{-0.3}^{+0.35}$ & $0.27_{-0.09}^{+0.08}$ \\
$1.56 - 3.13$ & $1.00 - 1.41$ & $0.46_{-0.1}^{+0.09}$ & $0.25_{-0.09}^{+0.11}$ & $0.54_{-0.17}^{+0.19}$ & $1.04_{-0.37}^{+0.41}$ & $0.46_{-0.09}^{+0.1}$ \\
$1.56 - 3.13$ & $1.41 - 2.00$ & $0.52_{-0.09}^{+0.1}$ & $0.14_{-0.07}^{+0.08}$ & $0.78_{-0.16}^{+0.2}$ & $1.2_{-0.36}^{+0.4}$ & $0.52_{-0.1}^{+0.1}$ \\
$1.56 - 3.13$ & $2.00 - 2.83$ & $0.1_{-0.05}^{+0.05}$ & $0.04_{-0.03}^{+0.04}$ & $0.13_{-0.07}^{+0.1}$ & $0.35_{-0.18}^{+0.25}$ & $0.1_{-0.05}^{+0.06}$ \\
$1.56 - 3.13$ & $2.83 - 4.00$ & $0.05_{-0.03}^{+0.03}$ & $0.03_{-0.02}^{+0.03}$ & $0.07_{-0.05}^{+0.06}$ & $0.18_{-0.11}^{+0.15}$ & $0.05_{-0.03}^{+0.03}$ \\
$1.56 - 3.13$ & $4.00 - 5.66$ & $0.02_{-0.01}^{+0.02}$ & $<0.04$ & $0.05_{-0.03}^{+0.05}$ & $<0.17$ & $0.02_{-0.01}^{+0.02}$ \\
$1.56 - 3.13$ & $5.66 - 8.00$ & $<0.02$ & $<0.04$ & $<0.05$ & $<0.13$ & $<0.02$ \\
$1.56 - 3.13$ & $8.00 - 11.31$ & $0.01_{-0.01}^{+0.02}$ & $<0.05$ & $0.03_{-0.02}^{+0.03}$ & $<0.16$ & $0.01_{-0.01}^{+0.01}$ \\
$1.56 - 3.13$ & $11.31 - 16.00$ & $0.09_{-0.03}^{+0.04}$ & $0.06_{-0.03}^{+0.05}$ & $0.16_{-0.07}^{+0.09}$ & $0.12_{-0.07}^{+0.11}$ & $0.09_{-0.03}^{+0.04}$ \\
\hline
$3.13 - 6.25$ & $0.50 - 0.71$ & $0.34_{-0.18}^{+0.18}$ & $0.11_{-0.07}^{+0.1}$ & $0.49_{-0.26}^{+0.33}$ & $0.72_{-0.39}^{+0.51}$ & $0.34_{-0.16}^{+0.19}$ \\
$3.13 - 6.25$ & $0.71 - 1.00$ & $0.71_{-0.21}^{+0.21}$ & $0.27_{-0.14}^{+0.16}$ & $1.04_{-0.34}^{+0.38}$ & $1.19_{-0.55}^{+0.67}$ & $0.69_{-0.19}^{+0.2}$ \\
$3.13 - 6.25$ & $1.00 - 1.41$ & $1.37_{-0.23}^{+0.23}$ & $0.86_{-0.23}^{+0.26}$ & $1.58_{-0.38}^{+0.47}$ & $2.67_{-0.72}^{+0.92}$ & $1.39_{-0.23}^{+0.23}$ \\
$3.13 - 6.25$ & $1.41 - 2.00$ & $1.79_{-0.26}^{+0.27}$ & $0.54_{-0.18}^{+0.2}$ & $2.56_{-0.43}^{+0.46}$ & $4.36_{-0.94}^{+1.0}$ & $1.82_{-0.24}^{+0.25}$ \\
$3.13 - 6.25$ & $2.00 - 2.83$ & $0.99_{-0.18}^{+0.18}$ & $0.4_{-0.15}^{+0.18}$ & $1.14_{-0.3}^{+0.35}$ & $2.8_{-0.7}^{+0.81}$ & $0.99_{-0.17}^{+0.18}$ \\
$3.13 - 6.25$ & $2.83 - 4.00$ & $0.27_{-0.1}^{+0.11}$ & $0.19_{-0.1}^{+0.13}$ & $0.3_{-0.16}^{+0.21}$ & $0.59_{-0.34}^{+0.35}$ & $0.27_{-0.1}^{+0.13}$ \\
$3.13 - 6.25$ & $4.00 - 5.66$ & $0.18_{-0.07}^{+0.09}$ & $0.13_{-0.07}^{+0.1}$ & $0.23_{-0.1}^{+0.13}$ & $0.29_{-0.19}^{+0.25}$ & $0.17_{-0.06}^{+0.08}$ \\
$3.13 - 6.25$ & $5.66 - 8.00$ & $0.08_{-0.05}^{+0.06}$ & $0.06_{-0.04}^{+0.05}$ & $0.1_{-0.06}^{+0.09}$ & $0.24_{-0.15}^{+0.23}$ & $0.08_{-0.04}^{+0.06}$ \\
$3.13 - 6.25$ & $8.00 - 11.31$ & $0.09_{-0.04}^{+0.06}$ & $<0.08$ & $0.18_{-0.09}^{+0.11}$ & $0.22_{-0.12}^{+0.19}$ & $0.09_{-0.05}^{+0.05}$ \\
$3.13 - 6.25$ & $11.31 - 16.00$ & $0.19_{-0.06}^{+0.08}$ & $0.13_{-0.06}^{+0.08}$ & $0.34_{-0.13}^{+0.15}$ & $<0.22$ & $0.19_{-0.06}^{+0.07}$ \\
\hline
$6.25 - 12.5$ & $0.50 - 0.71$ & $0.18_{-0.12}^{+0.17}$ & $<0.33$ & $0.39_{-0.26}^{+0.37}$ & $0.47_{-0.33}^{+0.49}$ & $0.19_{-0.13}^{+0.16}$ \\
$6.25 - 12.5$ & $0.71 - 1.00$ & $0.51_{-0.22}^{+0.26}$ & $0.2_{-0.14}^{+0.18}$ & $0.57_{-0.33}^{+0.4}$ & $1.38_{-0.75}^{+0.77}$ & $0.51_{-0.21}^{+0.24}$ \\
$6.25 - 12.5$ & $1.00 - 1.41$ & $2.34_{-0.39}^{+0.39}$ & $1.25_{-0.38}^{+0.4}$ & $3.12_{-0.68}^{+0.68}$ & $4.26_{-1.22}^{+1.36}$ & $2.32_{-0.35}^{+0.38}$ \\
$6.25 - 12.5$ & $1.41 - 2.00$ & $2.66_{-0.38}^{+0.4}$ & $1.58_{-0.44}^{+0.49}$ & $3.44_{-0.69}^{+0.68}$ & $4.95_{-1.43}^{+1.45}$ & $2.65_{-0.39}^{+0.37}$ \\
$6.25 - 12.5$ & $2.00 - 2.83$ & $3.37_{-0.39}^{+0.43}$ & $1.08_{-0.34}^{+0.39}$ & $4.12_{-0.69}^{+0.69}$ & $9.25_{-1.52}^{+1.65}$ & $3.37_{-0.39}^{+0.45}$ \\
$6.25 - 12.5$ & $2.83 - 4.00$ & $0.61_{-0.21}^{+0.22}$ & $0.49_{-0.2}^{+0.26}$ & $1.05_{-0.36}^{+0.42}$ & $0.41_{-0.28}^{+0.45}$ & $0.61_{-0.19}^{+0.21}$ \\
$6.25 - 12.5$ & $4.00 - 5.66$ & $0.17_{-0.09}^{+0.1}$ & $0.08_{-0.05}^{+0.08}$ & $0.34_{-0.18}^{+0.22}$ & $0.42_{-0.25}^{+0.36}$ & $0.17_{-0.08}^{+0.11}$ \\
$6.25 - 12.5$ & $5.66 - 8.00$ & $0.07_{-0.05}^{+0.06}$ & $0.09_{-0.06}^{+0.08}$ & $0.12_{-0.08}^{+0.13}$ & $<0.54$ & $0.07_{-0.04}^{+0.06}$ \\
$6.25 - 12.5$ & $8.00 - 11.31$ & $0.16_{-0.07}^{+0.09}$ & $0.1_{-0.07}^{+0.1}$ & $0.16_{-0.09}^{+0.15}$ & $0.46_{-0.25}^{+0.36}$ & $0.15_{-0.06}^{+0.08}$ \\
$6.25 - 12.5$ & $11.31 - 16.00$ & $0.18_{-0.07}^{+0.08}$ & $0.2_{-0.1}^{+0.12}$ & $0.26_{-0.13}^{+0.16}$ & $<0.49$ & $0.18_{-0.08}^{+0.09}$ \\
\hline
$12.5 - 25.0$ & $0.50 - 0.71$ & $<0.28$ & $<0.5$ & $<0.78$ & $<0.98$ & $<0.32$ \\
$12.5 - 25.0$ & $0.71 - 1.00$ & $0.26_{-0.16}^{+0.22}$ & $0.19_{-0.13}^{+0.29}$ & $0.66_{-0.4}^{+0.51}$ & $0.94_{-0.61}^{+0.86}$ & $0.28_{-0.18}^{+0.22}$ \\
$12.5 - 25.0$ & $1.00 - 1.41$ & $1.59_{-0.37}^{+0.44}$ & $1.0_{-0.48}^{+0.46}$ & $1.15_{-0.58}^{+0.72}$ & $3.84_{-1.56}^{+1.68}$ & $1.6_{-0.4}^{+0.49}$ \\
$12.5 - 25.0$ & $1.41 - 2.00$ & $2.41_{-0.51}^{+0.59}$ & $1.08_{-0.46}^{+0.51}$ & $3.18_{-0.89}^{+0.99}$ & $5.52_{-1.92}^{+2.07}$ & $2.42_{-0.48}^{+0.53}$ \\
$12.5 - 25.0$ & $2.00 - 2.83$ & $6.56_{-0.73}^{+0.69}$ & $2.82_{-0.6}^{+0.63}$ & $8.43_{-1.23}^{+1.16}$ & $15.39_{-2.28}^{+2.57}$ & $6.59_{-0.71}^{+0.71}$ \\
$12.5 - 25.0$ & $2.83 - 4.00$ & $2.13_{-0.46}^{+0.46}$ & $1.27_{-0.45}^{+0.49}$ & $2.89_{-0.74}^{+0.95}$ & $2.87_{-1.28}^{+1.54}$ & $2.1_{-0.39}^{+0.51}$ \\
$12.5 - 25.0$ & $4.00 - 5.66$ & $0.19_{-0.11}^{+0.17}$ & $0.26_{-0.16}^{+0.22}$ & $0.23_{-0.16}^{+0.28}$ & $0.76_{-0.49}^{+0.7}$ & $0.2_{-0.12}^{+0.16}$ \\
$12.5 - 25.0$ & $5.66 - 8.00$ & $0.24_{-0.11}^{+0.13}$ & $0.1_{-0.06}^{+0.11}$ & $0.34_{-0.19}^{+0.23}$ & $0.81_{-0.47}^{+0.54}$ & $0.24_{-0.11}^{+0.14}$ \\
$12.5 - 25.0$ & $8.00 - 11.31$ & $0.06_{-0.04}^{+0.07}$ & $<0.2$ & $0.16_{-0.11}^{+0.18}$ & $<0.63$ & $0.05_{-0.04}^{+0.06}$ \\
$12.5 - 25.0$ & $11.31 - 16.00$ & $0.18_{-0.09}^{+0.1}$ & $0.17_{-0.09}^{+0.13}$ & $0.26_{-0.15}^{+0.19}$ & $0.45_{-0.27}^{+0.41}$ & $0.17_{-0.08}^{+0.11}$ \\
\hline
$25.0 - 50.0$ & $0.50 - 0.71$ & $<0.57$ & $<0.93$ & $<1.25$ & $<2.49$ & $<0.61$ \\
$25.0 - 50.0$ & $0.71 - 1.00$ & $0.25_{-0.16}^{+0.26}$ & $<0.52$ & $0.34_{-0.24}^{+0.42}$ & $1.37_{-0.94}^{+1.28}$ & $0.27_{-0.17}^{+0.24}$ \\
$25.0 - 50.0$ & $1.00 - 1.41$ & $0.43_{-0.25}^{+0.38}$ & $0.34_{-0.22}^{+0.35}$ & $0.41_{-0.29}^{+0.53}$ & $2.34_{-1.37}^{+1.86}$ & $0.42_{-0.26}^{+0.39}$ \\
$25.0 - 50.0$ & $1.41 - 2.00$ & $2.19_{-0.55}^{+0.64}$ & $0.73_{-0.4}^{+0.54}$ & $2.47_{-0.98}^{+1.1}$ & $6.48_{-2.36}^{+2.69}$ & $2.18_{-0.55}^{+0.61}$ \\
$25.0 - 50.0$ & $2.00 - 2.83$ & $7.57_{-0.88}^{+0.98}$ & $4.25_{-0.84}^{+0.91}$ & $10.48_{-1.82}^{+1.82}$ & $12.55_{-3.01}^{+3.1}$ & $7.55_{-0.81}^{+0.87}$ \\
$25.0 - 50.0$ & $2.83 - 4.00$ & $3.45_{-0.63}^{+0.65}$ & $1.4_{-0.56}^{+0.62}$ & $5.94_{-1.36}^{+1.51}$ & $3.31_{-1.53}^{+2.04}$ & $3.41_{-0.64}^{+0.66}$ \\
$25.0 - 50.0$ & $4.00 - 5.66$ & $0.26_{-0.17}^{+0.22}$ & $0.52_{-0.29}^{+0.38}$ & $0.39_{-0.26}^{+0.42}$ & $0.74_{-0.53}^{+0.82}$ & $0.26_{-0.17}^{+0.21}$ \\
$25.0 - 50.0$ & $5.66 - 8.00$ & $0.27_{-0.13}^{+0.16}$ & $0.26_{-0.16}^{+0.24}$ & $0.22_{-0.15}^{+0.25}$ & $1.02_{-0.64}^{+0.86}$ & $0.28_{-0.14}^{+0.16}$ \\
$25.0 - 50.0$ & $8.00 - 11.31$ & $0.2_{-0.12}^{+0.14}$ & $<0.32$ & $0.34_{-0.21}^{+0.31}$ & $1.01_{-0.66}^{+1.18}$ & $0.2_{-0.12}^{+0.15}$ \\
$25.0 - 50.0$ & $11.31 - 16.00$ & $0.3_{-0.13}^{+0.18}$ & $0.27_{-0.15}^{+0.21}$ & $0.49_{-0.24}^{+0.35}$ & $<1.21$ & $0.29_{-0.12}^{+0.15}$ \\
\hline
$50.0 - 100.0$ & $0.50 - 0.71$ & $<1.11$ & $<2.93$ & $<2.87$ & $<5.3$ & $<1.05$ \\
$50.0 - 100.0$ & $0.71 - 1.00$ & $0.28_{-0.2}^{+0.33}$ & $<1.05$ & $<1.32$ & $1.39_{-0.95}^{+1.66}$ & $0.24_{-0.17}^{+0.28}$ \\
$50.0 - 100.0$ & $1.00 - 1.41$ & $0.33_{-0.23}^{+0.36}$ & $0.4_{-0.27}^{+0.45}$ & $0.57_{-0.4}^{+0.67}$ & $1.61_{-1.12}^{+1.97}$ & $0.36_{-0.25}^{+0.36}$ \\
$50.0 - 100.0$ & $1.41 - 2.00$ & $1.5_{-0.64}^{+0.72}$ & $0.6_{-0.39}^{+0.51}$ & $1.2_{-0.77}^{+0.89}$ & $6.45_{-2.61}^{+3.34}$ & $1.53_{-0.59}^{+0.58}$ \\
$50.0 - 100.0$ & $2.00 - 2.83$ & $6.52_{-1.03}^{+1.18}$ & $2.96_{-0.89}^{+1.06}$ & $10.2_{-2.04}^{+1.94}$ & $10.09_{-3.36}^{+4.83}$ & $6.6_{-1.08}^{+1.22}$ \\
$50.0 - 100.0$ & $2.83 - 4.00$ & $3.62_{-0.81}^{+0.89}$ & $2.33_{-0.84}^{+1.08}$ & $4.56_{-1.55}^{+1.65}$ & $5.28_{-2.45}^{+2.94}$ & $3.69_{-0.86}^{+0.94}$ \\
$50.0 - 100.0$ & $4.00 - 5.66$ & $0.92_{-0.41}^{+0.49}$ & $1.04_{-0.51}^{+0.68}$ & $1.47_{-0.73}^{+0.9}$ & $<2.79$ & $0.9_{-0.41}^{+0.43}$ \\
$50.0 - 100.0$ & $5.66 - 8.00$ & $0.53_{-0.26}^{+0.33}$ & $0.56_{-0.32}^{+0.54}$ & $0.67_{-0.41}^{+0.55}$ & $1.18_{-0.79}^{+1.28}$ & $0.51_{-0.25}^{+0.34}$ \\
$50.0 - 100.0$ & $8.00 - 11.31$ & $0.58_{-0.26}^{+0.33}$ & $0.54_{-0.3}^{+0.41}$ & $0.53_{-0.33}^{+0.48}$ & $1.5_{-0.91}^{+1.23}$ & $0.6_{-0.28}^{+0.31}$ \\
$50.0 - 100.0$ & $11.31 - 16.00$ & $0.37_{-0.21}^{+0.26}$ & $0.46_{-0.25}^{+0.36}$ & $0.58_{-0.33}^{+0.49}$ & $<2.12$ & $0.37_{-0.19}^{+0.27}$ \\
\hline
$100.0 - 200.0$ & $0.50 - 0.71$ & $<3.46$ & $<9.11$ & $<8.1$ & $<10.23$ & $<4.34$ \\
$100.0 - 200.0$ & $0.71 - 1.00$ & $<1.2$ & $<2.19$ & $<2.72$ & $<4.38$ & $<1.59$ \\
$100.0 - 200.0$ & $1.00 - 1.41$ & $0.41_{-0.28}^{+0.48}$ & $<1.25$ & $<1.79$ & $2.65_{-1.71}^{+2.55}$ & $0.52_{-0.36}^{+0.64}$ \\
$100.0 - 200.0$ & $1.41 - 2.00$ & $1.04_{-0.61}^{+0.71}$ & $0.62_{-0.43}^{+0.79}$ & $1.48_{-0.88}^{+1.22}$ & $4.07_{-2.48}^{+3.46}$ & $1.03_{-0.58}^{+0.71}$ \\
$100.0 - 200.0$ & $2.00 - 2.83$ & $6.01_{-1.23}^{+1.32}$ & $2.96_{-1.2}^{+1.56}$ & $7.23_{-2.14}^{+2.67}$ & $13.12_{-4.89}^{+5.26}$ & $5.86_{-1.26}^{+1.39}$ \\
$100.0 - 200.0$ & $2.83 - 4.00$ & $2.82_{-0.97}^{+1.07}$ & $1.21_{-0.71}^{+1.1}$ & $4.83_{-1.92}^{+2.16}$ & $3.96_{-2.4}^{+2.73}$ & $2.81_{-0.87}^{+1.05}$ \\
$100.0 - 200.0$ & $4.00 - 5.66$ & $0.67_{-0.39}^{+0.56}$ & $<0.89$ & $1.79_{-1.04}^{+1.35}$ & $1.7_{-1.2}^{+2.0}$ & $0.66_{-0.43}^{+0.54}$ \\
$100.0 - 200.0$ & $5.66 - 8.00$ & $0.86_{-0.45}^{+0.51}$ & $0.63_{-0.38}^{+0.54}$ & $1.68_{-0.89}^{+1.11}$ & $<3.09$ & $0.91_{-0.45}^{+0.54}$ \\
$100.0 - 200.0$ & $8.00 - 11.31$ & $1.55_{-0.59}^{+0.69}$ & $0.71_{-0.44}^{+0.8}$ & $2.98_{-1.11}^{+1.49}$ & $1.64_{-1.1}^{+1.78}$ & $1.49_{-0.53}^{+0.74}$ \\
$100.0 - 200.0$ & $11.31 - 16.00$ & $0.81_{-0.38}^{+0.47}$ & $1.08_{-0.58}^{+0.72}$ & $0.88_{-0.55}^{+0.83}$ & $2.25_{-1.45}^{+1.94}$ & $0.82_{-0.41}^{+0.45}$ \\
\hline
$200.0 - 400.0$ & $0.50 - 0.71$ & $<14.11$ & $<42.06$ & $<33.44$ & $<41.64$ & $<13.07$ \\
$200.0 - 400.0$ & $0.71 - 1.00$ & $<3.37$ & $<8.52$ & $<8.3$ & $<11.45$ & $<2.72$ \\
$200.0 - 400.0$ & $1.00 - 1.41$ & $<1.43$ & $<2.81$ & $<3.58$ & $<6.8$ & $<1.44$ \\
$200.0 - 400.0$ & $1.41 - 2.00$ & $0.57_{-0.42}^{+0.68}$ & $<1.75$ & $<3.3$ & $3.74_{-2.44}^{+3.88}$ & $0.55_{-0.4}^{+0.73}$ \\
$200.0 - 400.0$ & $2.00 - 2.83$ & $4.78_{-1.68}^{+1.74}$ & $1.43_{-0.91}^{+1.42}$ & $8.48_{-3.19}^{+4.08}$ & $5.19_{-3.21}^{+4.97}$ & $4.81_{-1.63}^{+1.9}$ \\
$200.0 - 400.0$ & $2.83 - 4.00$ & $2.21_{-1.14}^{+1.41}$ & $1.6_{-0.97}^{+1.63}$ & $1.95_{-1.31}^{+2.24}$ & $10.01_{-5.47}^{+6.86}$ & $2.21_{-1.19}^{+1.33}$ \\
$200.0 - 400.0$ & $4.00 - 5.66$ & $1.31_{-0.79}^{+0.95}$ & $1.05_{-0.69}^{+1.22}$ & $2.42_{-1.44}^{+2.08}$ & $3.64_{-2.41}^{+3.77}$ & $1.24_{-0.73}^{+1.05}$ \\
$200.0 - 400.0$ & $5.66 - 8.00$ & $2.75_{-1.05}^{+1.24}$ & $1.48_{-0.9}^{+1.3}$ & $5.63_{-2.34}^{+2.56}$ & $<5.33$ & $2.79_{-1.14}^{+1.27}$ \\
$200.0 - 400.0$ & $8.00 - 11.31$ & $1.09_{-0.63}^{+0.87}$ & $0.99_{-0.65}^{+1.1}$ & $1.79_{-1.24}^{+1.59}$ & $3.9_{-2.67}^{+3.98}$ & $1.09_{-0.65}^{+0.84}$ \\
$200.0 - 400.0$ & $11.31 - 16.00$ & $1.66_{-0.78}^{+0.95}$ & $1.08_{-0.67}^{+1.14}$ & $2.02_{-1.22}^{+1.79}$ & $6.52_{-3.6}^{+5.68}$ & $1.66_{-0.78}^{+0.87}$ \\
\end{longtable*}

\section{ExoPAG SAG13 Recommended Grids}\label{sec:exopag}

Table \ref{tbl:exopag} gives our occurrence rate results for FGK-, F-, G-, and K-type stars over the ExoPAG SAG13 recommended period-radius grid. The median and 68.3\% credible interval of each ABC posterior is shown. For bins with zero planet detections, we report only the upper limit (84.1th percentile).

\begin{longtable*}[h!]{c|c|c|c|c}
\caption{Occurrence rate results for F, G, and K type stars over the ExoPAG SAG13 recommended period-radius grid. Results are given in \% ($10^{-2}$).}\label{tbl:exopag}\\
\hline\hline
Period (days) & Radius ($R_{\bigoplus}$) & F (\%) & G (\%) & K (\%) \\
\hline
\endfirsthead
Period (days) & Radius ($R_{\bigoplus}$) & F (\%) & G (\%) & K (\%) \\
\hline
\endhead
$10.0 - 20.0$ & $0.67 - 1.00$ & $0.22_{-0.16}^{+0.26}$ & $0.5_{-0.31}^{+0.39}$ & $0.76_{-0.52}^{+0.85}$ \\
$10.0 - 20.0$ & $1.00 - 1.50$ & $1.83_{-0.58}^{+0.66}$ & $2.72_{-0.71}^{+0.7}$ & $4.8_{-1.49}^{+1.53}$ \\
$10.0 - 20.0$ & $1.50 - 2.25$ & $1.83_{-0.52}^{+0.6}$ & $6.21_{-0.99}^{+1.13}$ & $10.75_{-2.19}^{+2.47}$ \\
$10.0 - 20.0$ & $2.25 - 3.38$ & $2.19_{-0.51}^{+0.52}$ & $6.4_{-0.84}^{+1.08}$ & $9.32_{-1.85}^{+2.02}$ \\
$10.0 - 20.0$ & $3.38 - 5.06$ & $0.35_{-0.19}^{+0.24}$ & $1.09_{-0.44}^{+0.5}$ & $0.89_{-0.54}^{+0.76}$ \\
$10.0 - 20.0$ & $5.06 - 7.59$ & $0.34_{-0.18}^{+0.24}$ & $1.07_{-0.43}^{+0.45}$ & $0.86_{-0.54}^{+0.72}$ \\
$10.0 - 20.0$ & $7.59 - 11.39$ & $0.17_{-0.1}^{+0.14}$ & $0.27_{-0.16}^{+0.23}$ & $0.52_{-0.32}^{+0.48}$ \\
$10.0 - 20.0$ & $11.39 - 17.09$ & $0.1_{-0.06}^{+0.1}$ & $0.25_{-0.15}^{+0.2}$ & $0.55_{-0.33}^{+0.46}$ \\
\hline
$20.0 - 40.0$ & $0.67 - 1.00$ & $0.22_{-0.16}^{+0.26}$ & $0.44_{-0.27}^{+0.49}$ & $1.4_{-0.91}^{+1.45}$ \\
$20.0 - 40.0$ & $1.00 - 1.50$ & $0.46_{-0.25}^{+0.36}$ & $1.08_{-0.57}^{+0.68}$ & $5.05_{-1.89}^{+2.35}$ \\
$20.0 - 40.0$ & $1.50 - 2.25$ & $1.78_{-0.6}^{+0.64}$ & $3.58_{-1.02}^{+1.19}$ & $9.13_{-2.59}^{+3.08}$ \\
$20.0 - 40.0$ & $2.25 - 3.38$ & $4.81_{-0.77}^{+0.82}$ & $10.89_{-1.55}^{+1.68}$ & $10.18_{-2.18}^{+2.52}$ \\
$20.0 - 40.0$ & $3.38 - 5.06$ & $0.53_{-0.31}^{+0.43}$ & $0.97_{-0.53}^{+0.61}$ & $1.24_{-0.81}^{+1.19}$ \\
$20.0 - 40.0$ & $5.06 - 7.59$ & $0.57_{-0.33}^{+0.45}$ & $<1.59$ & $1.15_{-0.74}^{+1.16}$ \\
$20.0 - 40.0$ & $7.59 - 11.39$ & $<0.69$ & $0.16_{-0.12}^{+0.24}$ & $1.14_{-0.66}^{+0.98}$ \\
$20.0 - 40.0$ & $11.39 - 17.09$ & $0.13_{-0.09}^{+0.23}$ & $0.37_{-0.22}^{+0.27}$ & $<1.49$ \\
\hline
$40.0 - 80.0$ & $0.67 - 1.00$ & $<0.84$ & $<1.07$ & $1.32_{-0.95}^{+1.69}$ \\
$40.0 - 80.0$ & $1.00 - 1.50$ & $0.29_{-0.2}^{+0.31}$ & $0.61_{-0.41}^{+0.61}$ & $3.13_{-1.73}^{+2.28}$ \\
$40.0 - 80.0$ & $1.50 - 2.25$ & $0.56_{-0.34}^{+0.49}$ & $4.64_{-1.37}^{+1.59}$ & $10.08_{-3.23}^{+3.63}$ \\
$40.0 - 80.0$ & $2.25 - 3.38$ & $3.78_{-0.95}^{+0.96}$ & $12.29_{-1.9}^{+2.08}$ & $10.65_{-3.17}^{+3.71}$ \\
$40.0 - 80.0$ & $3.38 - 5.06$ & $1.22_{-0.6}^{+0.73}$ & $1.74_{-0.85}^{+0.96}$ & $0.95_{-0.67}^{+1.18}$ \\
$40.0 - 80.0$ & $5.06 - 7.59$ & $1.23_{-0.63}^{+0.78}$ & $1.73_{-0.83}^{+1.05}$ & $<2.2$ \\
$40.0 - 80.0$ & $7.59 - 11.39$ & $0.85_{-0.41}^{+0.52}$ & $0.48_{-0.31}^{+0.46}$ & $0.78_{-0.58}^{+1.09}$ \\
$40.0 - 80.0$ & $11.39 - 17.09$ & $0.24_{-0.16}^{+0.25}$ & $0.47_{-0.3}^{+0.47}$ & $<2.45$ \\
\hline
$80.0 - 160.0$ & $0.67 - 1.00$ & $<2.03$ & $<2.77$ & $<4.34$ \\
$80.0 - 160.0$ & $1.00 - 1.50$ & $<1.21$ & $0.86_{-0.56}^{+1.09}$ & $2.01_{-1.28}^{+1.83}$ \\
$80.0 - 160.0$ & $1.50 - 2.25$ & $1.13_{-0.71}^{+0.94}$ & $3.74_{-1.65}^{+1.79}$ & $7.46_{-3.24}^{+4.31}$ \\
$80.0 - 160.0$ & $2.25 - 3.38$ & $4.25_{-1.26}^{+1.64}$ & $10.87_{-2.31}^{+2.57}$ & $11.44_{-3.91}^{+4.45}$ \\
$80.0 - 160.0$ & $3.38 - 5.06$ & $0.67_{-0.45}^{+0.68}$ & $2.74_{-1.2}^{+1.51}$ & $1.85_{-1.25}^{+2.03}$ \\
$80.0 - 160.0$ & $5.06 - 7.59$ & $0.6_{-0.4}^{+0.56}$ & $2.65_{-1.16}^{+1.51}$ & $2.08_{-1.47}^{+2.12}$ \\
$80.0 - 160.0$ & $7.59 - 11.39$ & $0.96_{-0.5}^{+0.68}$ & $0.87_{-0.57}^{+0.79}$ & $<2.97$ \\
$80.0 - 160.0$ & $11.39 - 17.09$ & $0.99_{-0.51}^{+0.69}$ & $2.29_{-1.0}^{+1.2}$ & $1.11_{-0.79}^{+1.49}$ \\
\hline
$160.0 - 320.0$ & $0.67 - 1.00$ & $<6.03$ & $<7.67$ & $<12.51$ \\
$160.0 - 320.0$ & $1.00 - 1.50$ & $<2.08$ & $1.78_{-1.17}^{+1.8}$ & $3.62_{-2.43}^{+4.17}$ \\
$160.0 - 320.0$ & $1.50 - 2.25$ & $<1.73$ & $2.44_{-1.43}^{+2.17}$ & $4.95_{-3.13}^{+5.24}$ \\
$160.0 - 320.0$ & $2.25 - 3.38$ & $2.54_{-1.22}^{+1.58}$ & $9.26_{-2.79}^{+3.02}$ & $17.12_{-6.86}^{+9.42}$ \\
$160.0 - 320.0$ & $3.38 - 5.06$ & $1.15_{-0.77}^{+1.07}$ & $2.27_{-1.37}^{+1.79}$ & $3.63_{-2.44}^{+4.11}$ \\
$160.0 - 320.0$ & $5.06 - 7.59$ & $1.1_{-0.73}^{+1.1}$ & $2.3_{-1.36}^{+1.93}$ & $<7.18$ \\
$160.0 - 320.0$ & $7.59 - 11.39$ & $1.47_{-0.8}^{+1.1}$ & $3.46_{-1.7}^{+2.11}$ & $2.1_{-1.52}^{+3.55}$ \\
$160.0 - 320.0$ & $11.39 - 17.09$ & $0.77_{-0.51}^{+0.85}$ & $3.65_{-1.69}^{+1.95}$ & $<7.77$ \\
\hline
$320.0 - 640.0$ & $0.67 - 1.00$ & $<45.0$ & $<34.14$ & $<50.7$ \\
$320.0 - 640.0$ & $1.00 - 1.50$ & $<10.9$ & $<11.68$ & $<24.36$ \\
$320.0 - 640.0$ & $1.50 - 2.25$ & $<6.61$ & $4.72_{-3.13}^{+5.15}$ & $<20.17$ \\
$320.0 - 640.0$ & $2.25 - 3.38$ & $<5.24$ & $6.57_{-4.1}^{+6.21}$ & $19.06_{-11.29}^{+15.63}$ \\
$320.0 - 640.0$ & $3.38 - 5.06$ & $<6.82$ & $6.88_{-4.3}^{+5.54}$ & $13.62_{-8.71}^{+14.93}$ \\
$320.0 - 640.0$ & $5.06 - 7.59$ & $2.99_{-2.19}^{+4.35}$ & $6.87_{-4.52}^{+5.49}$ & $<29.49$ \\
$320.0 - 640.0$ & $7.59 - 11.39$ & $<7.23$ & $7.1_{-4.56}^{+6.29}$ & $<17.87$ \\
$320.0 - 640.0$ & $11.39 - 17.09$ & $<6.48$ & $<11.45$ & $7.07_{-5.21}^{+9.97}$ \\
\end{longtable*}

\end{document}